\documentclass[a4paper,twocolumn,10pt,aps,prd,superscriptaddress,showpacs,showkeys,amsmath,amssymb,floatfix,nofootinbib]{revtex4-1}
\usepackage{lmodern}

\usepackage[T1]{fontenc}
\usepackage[utf8]{inputenc}
\usepackage{color}
\usepackage[unicode=true,pdfusetitle,
 bookmarks=true,bookmarksnumbered=true,bookmarksopen=true,bookmarksopenlevel=1,
 breaklinks=false,pdfborder={0 0 0},backref=false,colorlinks=true]
 {hyperref}
\hypersetup{
 citecolor=blue,filecolor=blue,linkcolor=blue,urlcolor=blue}
\usepackage[normalem]{ulem}
\makeatletter

\pdfpageheight\paperheight
\pdfpagewidth\paperwidth

 \@ifundefined{textcolor}{}
 {%
   \definecolor{BLACK}{gray}{0}
   \definecolor{WHITE}{gray}{1}
   \definecolor{RED}{rgb}{1,0,0}
   \definecolor{green}{rgb}{0,0.7,0}
   \definecolor{BLUE}{rgb}{0,0,1}
   \definecolor{CYAN}{cmyk}{1,0,0,0}
   \definecolor{MAGENTA}{cmyk}{0,1,0,0}
   \definecolor{YELLOW}{cmyk}{0,0,1,0}
 }

\renewcommand{\[}{\begin{equation}}
\renewcommand{\]}{\end{equation}} 
\usepackage{array}
\def\be{\begin{equation}}
\def\ee{\end{equation}}

\newcommand{\ak}{\alpha_\text{K}}
\newcommand{\ab}{\alpha_\text{B}}
\newcommand{\am}{\alpha_\text{M}}
\newcommand{\at}{\alpha_\text{T}}
\newcommand{\ah}{\alpha_\text{H}}
\newcommand{\cH}{\mathcal{H}}
\newcommand{\omp}{\Omega_{\mathcal{P}}}
\newcommand{\omm}{\Omega_{\text{m}}}

\makeatother

\begin{document}

\title{Non-standard gravitational waves imply gravitational slip: on the difficulty of partially hiding new gravitational degrees of freedom}
\author{Ignacy Sawicki}
\affiliation{Départment de Physique Théorique and Center for Astroparticle Physics,
	Université de Genève, Quai E. Ansermet 24, 1211 Genève, Switzerland}
\affiliation{Central European Institute for Cosmology and Fundamental Physics, Fyzikální ustáv Akademie věd ČR, Na Slovance 2, 182 21 Praha 8, Czech Republic}

\author{Ippocratis D. Saltas}
\affiliation{Instituto de Astrofísica e Ciências do Espaço, Faculdade de Ciências, Campo Grande, PT1749-016 Lisboa, Portugal}

\author{Mariele Motta}
\affiliation{Départment de Physique Théorique and Center for Astroparticle Physics,
	Université de Genève, Quai E. Ansermet 24, 1211 Genève, Switzerland}
	
\author{Luca Amendola}
\affiliation{Intitut für Theoretische Physik, Ruprecht-Karls-Universität Heidelberg,
Philosophenweg 16, 69120 Heidelberg, Germany}

\author{Martin Kunz}
\affiliation{Départment de Physique Théorique and Center for Astroparticle Physics,
	Université de Genève, Quai E. Ansermet 24, 1211 Genève, Switzerland}

\begin{abstract}
 In many generalized models of gravity, perfect fluids in cosmology give rise to gravitational slip. Simultaneously, in very broad classes of such models, the propagation of gravitational waves is altered. We investigate the extent to which there is a one-to-one relationship between these two properties in three classes of models with one extra degree of freedom: scalar (Horndeski and beyond), vector (Einstein-Aether) and tensor (bimetric).
 
 We prove that in bimetric gravity and Einstein-Aether, it is impossible to dynamically hide the gravitational slip on all scales whenever the propagation of gravitational waves is modified. Horndeski models are much more flexible, but it is nonetheless only possible to hide gravitational slip dynamically when the action for perturbations is tuned to evolve in time toward a divergent kinetic term. These results provide an explicit, theoretical argument for the interpretation of future observations if they disfavoured the presence of gravitational slip.
\end{abstract}

\maketitle

\section{Introduction}\label{sec:Introduction}
	
The nature of the late-time acceleration of the Universe remains elusive. Although observations favor General Relativity (GR) with a cosmological constant, it is still far from clear whether the underlying model of gravity at large scales does not involve some other, dynamical degree of freedom or a genuine modification of GR. 

The distinction between the various classes of dark-energy models is one of the biggest challenges of future large-scale-structure surveys such as the Euclid satellite mission \cite{Amendola:2016saw}. It is well known that many models of dark energy exhibit non-vanishing gravitational slip in the presence of perfect-fluid matter sources. Moreover, it was shown in Refs~\cite{Amendola:2012ky,Motta:2013cwa} that the gravitational slip in the baryon Jordan frame can be expressed solely in terms of  quantities observable on the sky, and therefore is itself a \emph{bona fide} observable. Fisher matrix forecasts for a Euclid-like survey predict an accuracy of $10\%$ in the measurement of the quantity $\eta$ parameterizing the slip, if it is assumed to be scale independent~\cite{Amendola:2013qna}.

In Ref.~\cite{Saltas:2014dha}, we pointed out that models in which perfect-fluid matter generates gravitational slip at linear order in perturbations on cosmological backgrounds generically  also has a modified propagation of gravitational waves (GWs). Thus a detection of a non-zero slip at late times would in principle imply a genuine modification of GR. Understanding the connection between the physics of tensors and scalars has  become all the more pertinent since the detection of gravitational waves by the LIGO collaboration \cite{Abbott:2016blz}. Depending on the rate of progress in cosmological surveys and GW observations, one of these fields should be able to inform on the other.

The remaining question is to what extent this relationship is guaranteed i.e.\ \emph{whether the vanishing of gravitational slip always implies that gravitational waves propagate in a standard manner, or whether it is possible to hide the slip and dynamically maintain this under evolution in time}. In this article, we refer to the ability of the model to arrange its degrees of freedom at the linear level such that no gravitational slip is sourced by perfect-fluid matter as the \emph{dynamical shielding} of the gravitational slip. We investigate three very broad classes of modified gravity models for the existence of this property, (i) Horndeski scalar-tensor theories \cite{Horndeski:1974wa, Deffayet:2011gz}, (ii) Einstein-Aether vector-tensor models \cite{Jacobson:2000xp} and (iii) bimetric theories of massive gravity \cite{Hassan:2011zd}. 

For successful dynamical shielding of slip, we require two features:
\begin{enumerate}
	\item We first ask if it is at all possible to maintain a no-slip configuration in time, while GW propagation is modified. We prove that, in bimetric theories, linear gravitational slip cannot be shielded at all scales stably in time. On the other hand, in Horndeski and Einstein-Aether models, particular parameter choices do exists for which the shielding of gravitational slip is dynamically maintained. These models  correspond to models where the sound speed of the scalar degree of freedom vanishes and as such they are rather singular and peculiar limits of these theories, where the new fields do not propagate wave-like modes.
	 
	\item We then ask whether, for these particular choices of parameters, a no-slip configuration is actually an attractor of the dynamics, i.e.\ whether a generic initial conditions for perturbations will evolve to it. We find that in Einstein-Aether any slip present initially is maintained for all time. It is only in particular (and singular) Horndeski models that the initial slip decays away and then this no-slip condition can be maintained in time.
\end{enumerate}

This means that achieving the dynamical shielding of slip requires a scalar-tensor theory and a severe tuning of its parameters to a regime where the parameters of the action for perturbations diverge. Given this, if future observations conclude that gravitational slip is strongly disfavored, then it would be unlikely that gravity could be modified, in the sense of a modification of GW dynamics.

We also investigate the beyond-Horndeski models introduced in Refs~\cite{Zumalacarregui:2013pma,Gleyzes:2014dya,Gleyzes:2014qga} and show that they are an exception to our original assertion in Ref.~\cite{Saltas:2014dha}:  they seem to be the unique class of models where gravitational slip is generated from perfect-fluid matter without any modification in tensor propagation. The conclusion of this paper are summarized in the table below:

\begin{table}[h]
	\centering
	\begin{tabular}{lcccll}
		\hline\hline
		$\eta\neq 1$ 	&&$\Rightarrow$	&& mod.\ GW or BH & Section \ref{sec:AS>MG}\\
		\hline
		stan.\ GW 	&&$\Rightarrow$	&& $\eta=1$ or BH & Section \ref{sec:AS>MG} \\
		\hline
		$\eta=1$ 	&&$\Rightarrow$	&& stan.\ GW or tuned Horndeski & Section \ref{sec:MG>AS} \\
		\hline
		mod.\ GW 	&&$\Rightarrow$	&& $\eta\neq1$ or tuned Horndeski& Section \ref{sec:MG>AS} \\
		\hline\hline
	\end{tabular}
\caption{Summary of conclusions of this paper. Stan.\ --- standard, mod.\ --- modified, BH --- beyond Horndeski operators active. See relevant sections for details and assumptions. \label{tab:summary}}
\end{table}

We structure the paper as follows: In section \ref{sec:Assumptions}, we present the fundamental assumptions and notation we use in the rest of the paper. In section~\ref{sec:AS>MG}, we introduce the three classes of models on which we focus, Horndeski, Einstein-Aether and bimetric gravity and discuss how gravitational slip and modified tensor propagation emerge, reviewing the results of Ref.~\cite{Saltas:2014dha}. In section~\ref{sec:MG>AS}, we describe how to ensure that the shielding of gravitational slip be preserved under evolution in time and investigate the requirements on the model space this imposes. We draw our conclusions and discuss the implications of our analysis in section \ref{sec:Conclusions}.

\section{Assumptions and Definitions}\label{sec:Assumptions}

In this section, we lay down the main physical and mathematical assumptions we use throughout this paper.

We assume that the Universe is well described by small linear perturbations
living on a spatially flat Friedmann-Lemaître-Robertson-Walker (FLRW) metric. We take as the line element for the metric on which matter and light propagate in the Newtonian gauge:
\begin{align}
\mathrm{d}s^{2}=-(1+2\Psi)\mathrm{d}t^{2}+a^{2}(t)(1-2\Phi)\left[\delta_{ij}+h_{ij}\right]\mathrm{d}\boldsymbol{x}^{i}\mathrm{d}\boldsymbol{x}^{j}, \notag
\end{align}
where $\Phi(t,\boldsymbol{x})$ and $\Psi(t,\boldsymbol{x})$ are the scalar gravitational potentials in Newtonian gauge and $h_{ij}(t,\boldsymbol{x})$ is the transverse traceless spatial metric (tensor) perturbation, i.e.\ the gravitational wave of the Jordan-frame metric. The requirement of linearity requires that the scalar and tensor perturbations and their gradients all be small, 
\be
|\Phi|, |\Psi|, |h_{ij}| \ll1.
\ee 

From here on, we will express all the linear perturbation variables in momentum space with wavenumber $k$. Overdots and primes denote derivatives with respect to cosmic time and conformal time respectively, unless explicitly stated otherwise. $H$ is the Hubble parameter and $\mathcal{H}$ is the conformal Hubble parameter.
We assume that the matter sector behave as dust, neglecting the contributions from the neutrinos and radiation. This assumption is good enough for the late universe where the relativistic components are extremely subdominant.

The gravitational slip is defined as a difference between the two scalar potentials $\Phi$ and $\Psi$ and one typically defines
\[
\eta\equiv\frac{\Phi}{\Psi}\,,
\]
where in this equation the fields on the right-hand side are understood to be the root-mean-square values related to observables on the sky and therefore positive definite. In $\Lambda$CDM cosmology, there is no gravitational slip, $\eta=1$, with small corrections appearing from neutrino free-streaming. At second order in perturbations, gravitational slip also always appears even when the matter consists completely of dust \cite{Ballesteros:2011cm}, but in the late universe should be smaller than $\left|\eta-1\right|\lesssim10^{-3}$ \cite{Ballesteros:2011cm,Adamek:2013wja}.

The gravitational slip is generated through the anisotropy constraint, i.e.\ the traceless part of the $(ij)$ linearized Einstein equations. In modified-gravity models in Newtonian gauge, it takes the schematic form
\begin{equation}
C\equiv \Psi(t,k) - \Phi(t,k) = \sigma(t)\Pi(t,k) + \pi_\text{m}\,, \label{eq:Aniso0}
\end{equation}
with $\Pi(t,k)$ a functional of the linear perturbation variables of the model as well as potentially some background quantities; $\sigma(t)$ on the other hand is a background function only; $\pi_\text{m}$ is the amplitude of the scalar anisotropic stress of the matter components, which for the purpose of the proof in this paper we neglect. 

Notice that, to avoid ambiguity, we make a distinction between anisotropic stress and gravitational slip. We refer to the former as a property of matter and the latter as a property of the geometry. Thus the energy-momentum tensor of a fluid which exhibits anisotropic stress sources gravitational slip on the geometrical part of the linearized Einstein's equations. In modifications of gravity, gravitational slip can appear at linear order even without anisotropic stress.

In section~\ref{sec:AS>MG}, we connect the variables $\sigma$ and $\Pi$ to the particular models we will be interested in. For the moment, let us note that modified-gravity models do feature an $\mathcal{O}(1)$ correction to the slip parameter at linear order in perturbations, on at least some scales.

On the other hand, it is well known that the value of $\eta$ can be modified by a change of frame, e.g.\ a conformal rescaling
of the metric, making its value seemingly ambiguous.

In Refs~\cite{Amendola:2012ky,Motta:2013cwa}, it was shown that
comparing the evolution of redshift-space distortions of the galaxy
power spectrum with weak lensing tomography allows us to reconstruct
$\eta$ as a function of time and scale in a model-independent manner.
Such an operational definition removes the frame ambiguity since
the measurement picks out the particular metric on the geodesics of
which the galaxies and light move. It is the gravitational slip in
that metric that is being measured by such cosmological probes. With
the ambiguity of frame removed, the gravitational slip is a \emph{bona fide} observable, rather than just a phenomenological parameter. Fixing
the metric also determines what is considered a gravitational wave:
we call these the propagating spin-2 perturbations of the metric
on which baryonic matter moves.

Dynamical models of late-time acceleration can feature interactions between the new degree of freedom and the matter metric. As we have shown in Ref.~\cite{Saltas:2014dha}, these interactions lead to modifications in the propagation of tensor modes (gravitational waves). Depending on the model, on the FLRW background, \emph{the speed of tensor modes ($c_\text{\emph{T}}$) can be altered, the effective Planck mass $M_*$ can evolve in time or a graviton mass $\mu$ can appear, while the presence of a possible second tensor field sources the evolution through the term $\Gamma \gamma_{ij}$}, giving
for the tensor equation of motion
\begin{equation}
	h''_{ij}+ (2+\nu)\mathcal{H}h_{ij}'+c_\text{T}^2k^2 h_{ij} + a^2 \mu^2 h_{ij}=a^2\Gamma \gamma_{ij}\,. \label{eq:GWprop}
\end{equation}
The deviations away from the standard behavior are contained in  $\nu$, $c_\text{T}$ and $\mu$, with all these quantities defined in the Jordan frame of the matter and in principle free to be functions of time. As we will review in section~\ref{sec:AS>MG}, for any of these modifications to be present, $\sigma$ in Eq.~\eqref{eq:Aniso0} must also be non-zero. Thus gravitational slip sourced from perfect-fluid matter is a sign that gravitational-wave propagation is modified.

The question we aim to answer is whether it is possible to find model parameters where the degrees of freedom arrange themselves in such a way that the gravitational slip be hidden. We refer to this as the \emph{dynamical shielding} of the slip. We purposefully do not use the more usual term \emph{screening}, which we restrict to mean the hiding of modifications of gravity (including the slip) through non-linear effects in the dynamics (e.g.\ chameleon \cite{Khoury:2003aq} or Vainshtein screening \cite{Vainshtein:1972sx}). Screening implies that the scalar adopts a non-linear profile and therefore is a change of the background solution. As such, non-linear screening would also result in suppressing the modification to GW propagation.

\section{Gravitational Slip Implies Modified Tensors\label{sec:AS>MG}}

In Ref.~\cite{Saltas:2014dha}, we showed that for very general classes of modified gravity theories featuring one extra degree of freedom, whenever perfect-fluid matter sources gravitational slip,  the propagation of tensor modes is also modified. In this section, we shall briefly introduce the main properties of the three classes of models and review the appearance of gravitational slip and modifications of the propagation of GWs. We will focus on general classes of models which introduce one new scalar, vector or  tensor degree of freedom, respectively. 

\subsection{General modifications with an extra scalar: The Horndeski scalar-tensor model and beyond}\label{sec:Horndeski}

The Horndeski class of models encompasses all theories which can be constructed from the metric $g_{\mu\nu}$ and a single scalar field $\phi$ and which have equations of motion with no more than second derivatives. It includes the majority of the popular models of late-time acceleration such as quintessence, perfect fluids, Brans-Dicke gravity, $f(R)$ gravity, $f(G)$ gravity, kinetic gravity braiding and galileons. We refer to the scalar as the dark energy
(DE). The Horndeski Lagrangian is defined as the sum of four terms
$\mathcal{L}_{2}$ to $\mathcal{L}_{5}$ that are fully specified
by a non-canonical kinetic term $K(\phi,X)$ and three, in
principle arbitrary, coupling functions $G_{3,4,5}(\phi,X)$, where
$X=-g_{\mu\nu}\phi^{,\mu}\phi^{,\nu}/2$ is the canonical kinetic
energy term for the scalar field and where the comma denotes a partial derivative.

For Horndeski, we make extensive use of the formulation for linear structure formation
in scalar-tensor theories introduced in Ref.~\cite{Bellini:2014fua}, similar to the effective field theory approach of Refs~\cite{Gubitosi:2012hu,Bloomfield:2012ff,Gleyzes:2013ooa}.
The form of linear perturbation equations for all Horndeski models
can be completely described in terms of the background expansion history,
density fraction of matter today $\Omega_{\text{m}0}$, and four dimensionless,
independent and arbitrary functions of time only, $\alpha_{\text{K}},\alpha_{\text{B}},\alpha_{\text{M}}$
and $\alpha_{\text{T}}$, which mix the four free functions $K$ and $G_{i}$ present in the action. As we review below,
the parameters $\alpha_{\text{M}}$ and $\alpha_{\text{T}}$ both
affect the propagation of gravitational waves and control the appearance
of gravitational slip. The \emph{braiding} $\alpha_{\text{B}}$ controls
whether dark energy clusters at all, while the \emph{kineticity} $\alpha_{\text{K}}$
in turn controls at what scales this happens and acts to suppress
the sound speed of the scalar modes.

In Ref.~\cite{Bellini:2014fua}, it was shown that the anisotropy constraint in the notation of Eq.~\eqref{eq:Aniso0} is
\begin{align}
\sigma &= \am - \at \label{eq:anisoeq}\,,\\
\Pi &= \frac{\at}{\sigma}\Phi + H v_X \,,  \notag
\end{align}
where $v_X\equiv-\delta\phi/\dot{\phi}$ is a perturbation of the scalar field.

On the other hand, the propagation of GWs is modified by terms
\begin{align}
&\nu = \alpha_\text{M}\,,	&&	c_\text{T}^2 = 1+\alpha_\text{T}\,, \\
&\mu^2=0\,,					&&  \Gamma=0 \,, \notag
\end{align}
in the notation of Eq.~\eqref{eq:GWprop}.

It is clear from Eq.~\eqref{eq:anisoeq} that when both $\alpha_{\text{M}}=\alpha_{\text{T}}=0$, $\Phi=\Psi$. However, this choice of the parameters also switches off all the modifications in Eq.~\eqref{eq:GWprop}. In the context of scalar-tensor models, a detection of gravitational slip sourced by perfect-fluid matter therefore is direct evidence that one or both of the parameters $\alpha_{\text{T}}$ and $\alpha_{\text{M}}$ are different from their concordance values of zero and that gravity is modified.

\subsubsection*{Beyond Horndeski}
Recently, an extension to Horndeski models was discovered~\cite{Zumalacarregui:2013pma, Gleyzes:2014dya, Gleyzes:2014qga}. It contains higher derivatives in the Einstein equations, but which all cancel in the equations of motion for the real dynamical modes. Thus the theory propagates no more degrees of freedom than Horndeski. It is currently understood that theories containing only the beyond-Horndeski terms are ghost-free. When other scalar Lagrangian terms are included, the theory continues to propagate just the one scalar degree of freedom, and thus can be ghost-free, provided that the total Lagrangian is related to a Horndeski one through a disformal transformation. When this is not the case, the second degree of freedom reappears (see Refs~\cite{Langlois:2015cwa, Crisostomi:2016tcp} for details). However, since we have fixed the frame by requiring that it be the Jordan frame of these baryons, these models do represent new phenomenology.

Linear fluctuations in these \emph{beyond-Horndeski} models can be brought into the Horndeski form through a disformal transformation of the metric, but only at the price of introducing a non-minimal coupling in the matter sector. Thus, in the Jordan frame of the matter and light, they do introduce new physics parameterized by a new variable $\alpha_\text{H}$. However, $\alpha_\text{H}$ does not enter the GW Eq.~\eqref{eq:GWprop}, implying that the propagation of tensors is not modified. 

On the other hand, the anisotropy equation \eqref{eq:Aniso0} does have new contributions,
\begin{align}
\sigma &= \am- \at\,, \label{eq:beyond-aniso}\\
\Pi &= \frac{\at}{\sigma}\Phi - \frac{\ah}{\sigma}\left(\Psi + \dot{v}_X\right) +H v_X \,. \notag
\end{align}

The $\alpha_\text{H}$ parameterizes a new type of mixing between the scalar and the metric. In terms of the effective-field-theory notation of Ref.~\cite{Gleyzes:2013ooa,Gleyzes:2014qga}, in beyond-Horndeski models a $\alpha_\text{H}\delta N \delta R$ appears in the quadratic effective action instead of $\alpha_\text{B}\delta N \delta K$ which appears in the case of \emph{braiding}. $\delta N$ is the fluctuation of the lapse, $\delta K$ --- of the extrinsic curvature and $\delta R$ --- of the intrinsic curvature of the spatial slice, all in unitary gauge.

Since both these terms are made up of products of scalars, neither of them can contribute to the graviton equation of motion. On the other hand, $\delta R$ contains second spatial derivatives of a scalar curvature fluctuation, $\partial^2\zeta$, and therefore it does contribute to the anisotropy constraint.

Thus beyond-Horndeski models are counter-examples to our assertion in Ref.~\cite{Saltas:2014dha}: in this class of models, gravitational slip does \emph{not} imply that the propagation of gravitational waves be modified. Thus for a dust-filled universe, standard GWs together with a detection of slip would imply that a beyond-Horndeski modification be present.

\subsection{General modifications with an extra vector: the Einstein-Aether vector-tensor model}
The Einstein-Aether (EA) model introduces a new, spin-1 propagating degree of freedom, denoted as $u^\mu$. 
The vector field spontaneously breaks Lorentz symmetry due to its non-trivial vacuum expectation value (v.e.v), introducing a preferred direction. The theory space of such models is strongly constrained in the Solar-System~\cite{Bonvin:2007ap} and even more stringently by observation of binary pulsars~\cite{Yagi:2013ava,Yagi:2013qpa}.
It should be noted that this type of models does not provide a mechanism for acceleration and therefore requires a cosmological constant. Below, unless otherwise stated, we will use the equations and notation of Ref.~\cite{Lim:2004js}.

The dynamics of the vector field are specified through the following Lagrangian:
\begin{align}
 \mathcal{L}_\text{v}=&-\beta_1 \nabla_\mu u^\nu \nabla^\mu u_\nu - \beta_2 (\nabla_\mu u^\mu)^2 - \notag \\
  &-\beta_3 \nabla_\mu u ^\nu \nabla_\nu u^\mu + \lambda(u^\alpha u_\alpha + m^2), \label{lagrangian:EA}
\end{align}
where the $\beta_i$'s are free, constant parameters, $\lambda$ is a Lagrange multiplier which enforces the non-trivial v.e.v of $u^\mu$ in the action, and $m^2$ is the value of the vector's v.e.v. Using a kinematic decomposition of the velocity gradients in \eqref{lagrangian:EA} one can see  that the coefficient of the shear of the vector field $u_\mu$ is $-(\beta_1 + \beta_3)$.%
\footnote{Note that, in full generality, in the Lagrangian~\eqref{lagrangian:EA} one should also include the term $\sim \nabla^{\alpha} u^{\beta} \nabla_{\beta} u_{\alpha}$, which in a kinematical decomposition will give rise to the twist. However, this term vanishes on hypersurface-orthogonal settings \cite{Jacobson:2013xta}, such as the cosmological background. It would thus not contribute in linear perturbation theory in cosmology and we neglect it.}

The extra dynamical degree of freedom at the linear level in perturbations is the fluctuation of the spatial components $v^{i}$ of the vector $u^{\mu}$, assuming a decomposition of the latter as $u^{\mu} = \bar{u}^\mu + v^\mu$, where $\bar{u}^\mu$ denotes the background field. 
Conveniently defining the divergence of the vector's spatial perturbation as 
\begin{equation}
\Theta \equiv \left( \frac{a}{m} \right) \partial_i v^i \,, \label{eq:ThetaDef}
\end{equation}
in Ref.~\cite{Lim:2004js}, it was shown that the anisotropy constraint Eq.~\eqref{eq:Aniso0} becomes
\begin{align}
 \Pi = \frac{\left( a^2 \Theta \right)'}{a^2 k^2}\,, \qquad \sigma = 16\pi G m^2 (\beta_1+\beta_3),
\end{align}
with the prime denoting a derivative with respect to conformal time. 
On the other hand, the parameters of the tensor equation \eqref{eq:GWprop} are
\begin{align}
	&\nu=0\,,\quad 		&& c_\text{T}^2=(1-\beta_1-\beta_3)^{-1} \\
	&\mu^2 =0\,,\quad	&&\Gamma=0\,. \notag
\end{align}
Notice that, the modification of $c_\text{T}$ appears through the coefficient $\beta_1 + \beta_3$. This corresponds to the link between gravitational slip and the propagation of tensors for an EA theory, in the presence of perfect-fluid matter.

\subsection{General modifications with an extra tensor: bimetric massive gravity model}

The Hassan-Rosen bimetric theory \cite{Hassan:2011zd} is a natural extension of de Rham-Gabadadze-Tolley massive gravity \cite{deRham:2010ik,deRham:2011rn} where the fixed reference metric is promoted to a dynamical one. Despite the kinetic term for the reference metric, the theory remains ghost-free and has the advantage of allowing consistent flat FLRW solutions. For a recent review on the subject we refer the reader to Refs~\cite{deRham:2014zqa,Schmidt-May:2015vnx}.

The model contains two dynamical metric tensors, $g_{\mu\nu}, f_{\mu\nu}$, and matter must be minimally coupled to only one of the metrics in order to guarantee the absence of a ghost degree of freedom. We choose $g_{\mu\nu}$ to be this metric.%
\footnote{It is in principle possible to couple to one other particular combination of metrics which will cause a ghost to appear above the strong-coupling scale of the theory \cite{deRham:2014fha}.} %
The kinetic terms for each of the metrics are standard Einstein-Hilbert and the action is augmented by an interaction potential $U$ of the very specific form,
\begin{equation}
S_U = M_\text{Pl}^2\int \mathrm{d}^{4}x \sqrt{-g} \sum_{n=0}^{4} \beta_{n} e_{n}
\label{eq:bigravAction}
\end{equation}
where $e_{n}$ are symmetric polynomials built out of the quantity $X^{\mu}_{\nu} \equiv \sqrt{g^{\mu\alpha} f_{\alpha\nu}}$. The five $\beta_n$ are the only free parameters of this model, defined here to be dimensionful, absorbing the mass scale of the potential. Since matter is coupled minimally to $g$, the geodesics are going to be those of $g$ and therefore the observable potentials and gravitational waves are going to be those of $g$.

In Ref.~\cite{Comelli:2012db} it was shown that when both the background metrics are flat FLRW, the anisotropy constraint associated with the metric $g$, takes the form \footnote{For a similar analysis see also Ref.~\cite{Solomon:2014dua}.}
\begin{align}
	\Pi = \Delta E\,,\qquad \sigma = a^2 r \bar{Z} \qquad 
\end{align}
in the notation of Eq.~\eqref{eq:Aniso0}. $\Delta E$ is a gauge invariant variable, defined as $\Delta E\equiv E_g-E_f$, where $E_{g,f}$ is the scalar coming from the transverse spatial perturbation of the metric; $r,\bar{Z}$ are background functions of time only, defined in section~\ref{sec:Bimetric} \footnote{In the notation of Ref.~\cite{Comelli:2012db}, $r \bar{Z}=f_1$.}. 

On the other hand, the equation of motion for gravitational waves of the $g$ metric in notation of \eqref{eq:GWprop} is:
\begin{align}
&\nu=0\,, && c_\text{T}^2 = 1\,,\label{eq:big_GWmods}\\
&\mu^2 = r \bar{Z}\,, &&\Gamma = -r \bar{Z}\,. \notag
\end{align}
Thus, in these models, if perfect-fluid matter sources gravitational slip, $r\bar{Z}\neq0$ and the propagation of GWs is also modified. Note that the source $\gamma_{ij}$ in this model are the tensor fluctuations of the metric $f_{\mu\nu}$ around its FLRW background. Thus the two tensors are coupled, which turns out to lead to an instability in the radiation-dominated universe \cite{Cusin:2014psa}. 


\section{Modified Tensors Imply Gravitational Slip\label{sec:MG>AS}}

Having reviewed our previous results, we now turn to the converse question, which will be the main focus of the current work: \\
\emph{If the propagation of GWs is modified in cosmology, does this mean that the scalar sector must produce gravitational slip for perfect-fluid sources?
Or do there exist choices of model parameters which allow for configurations of the degrees of freedom where the gravitational slip is dynamically shielded at \emph{all} scales and where this shielding is stable under evolution in time?}

Let us start by introducing the mathematical method we will be using. We study the scalar perturbations on an FLRW background in a universe containing dust and a modification of gravity. We assume that we are not at some future de-Sitter attractor of the expanding Universe, where the matter has already diluted away. This is relevant, since such de Sitter configurations asymptotically lose the matter degrees of freedom and the evolution of parameters ceases. The dynamics become much simpler since there no matter to collapse and source slip. No-slip conditions can then be satisfied, \cite{Linder:2014fna}. They do not represent the observable universe, however.

First, we eliminate spurious degrees of freedom and auxiliary variables, i.e.~fix any gauge freedom and solve all the constraints in the Einstein equations so that  only the real and independent dynamical variables remain. In all of the cases considered here, there are two independent scalar degrees of freedom, described by two second-order equations of motion: one corresponding to the dust, the other --- to the new scalar in the dark-energy model, either introduced explicitly, or arising from the helicity-0 polarization of the new massive vector or tensor. 

We can of course choose an arbitrary basis for the phase space variables, which together with their corresponding velocities form the phase-space vector, 
\be
\boldsymbol{X}(t;k) =  \{\psi_1, \dot{\psi}_1, \psi_2, \dot{\psi}_2\}, \; \; \; \psi_i = \psi_i(t;k)\,.
\ee 
$\boldsymbol{X}(t;k)$ fully characterizes the configuration of the scalar perturbations of the universe at any one time $t$. Given the anisotropy equation (\ref{eq:Aniso0}), we now require that the initial configuration at time $t_0$ be chosen in such a way that there be no gravitational slip, i.e.\ that the independent degrees of freedom are arranged to
satisfy
\begin{equation}
C[\boldsymbol{X}] = \boldsymbol{A}^0(t_0;k) \cdot \boldsymbol{X}(t_0;k) =0\,,\label{eq:NoSlip-general}
\end{equation}
where the $A^0_{i}$ are the compact notation for the coefficients of the coordinates $\psi_i$, $\dot{\psi}_i$. The $A^0_i$ are functions of the model parameters, and in principle they are all functions of time and scale. Their exact form is determined by the model. The product $\boldsymbol{A}^0(t;k) \cdot \boldsymbol{X} (t;k)$ should be understood as a usual matrix product between two state vectors: one in model space ($A^0_i$), the other in phase space ($X^i$). For the concrete cases described here, $C$ is the no-slip constraint \eqref{eq:Aniso0}.

In view of Eq.~\eqref{eq:NoSlip-general}, the condition that the gravitational slip vanishes is therefore a 3-dimensional subspace $C=0$ with orthogonal vector $\boldsymbol{A}^0$ and passing though the origin $\boldsymbol{X}=0$ in the 4-dimensional space of the possible configurations. The origin is a point in the phase space where no perturbations are present, i.e.\ the exact background cosmology.

It is of course always possible to tune the initial conditions to satisfy $C=0$. However, the equations of motion for the system generate an evolution of this configuration in the full phase space, and the question is whether it is possible to find a model
where that evolution would be restricted to the constrained space $C=0$. For this it is
necessary that all time derivatives of $C$, when evaluated on the
equations of motion, also vanish (i.e.~time derivatives of constraints do not generate new constraints).

Taking time derivatives of $C$ and evaluating them on equations of motion generates  new constraints
\begin{equation}
\frac{\mathrm{d}^n{C}}{\mathrm{d}t^n} = \boldsymbol{A}^n\cdot \boldsymbol{X} \approx 0\,,
\end{equation}
where we use $\approx$ to signify equality on the equations of motion. Each of the constraints is a three-dimensional hyperplane passing through the origin with the associated orthogonal vector $\boldsymbol{A}^n$. Since the phase space is four-dimensional, there can at most be four such constraints, $C,\dot{C}, \ddot{C}, \dddot{C}$ with their four associated vectors which span the phase space.

If the four vectors $\boldsymbol{A}^n$ are all linearly independent, then the only intersection of all the hyperplanes is the origin, i.e.\ the only configuration for which the vanishing slip is maintained under time evolution is the configuration with no fluctuations. This is not of interest, but is the generic case. 

Nonetheless, one might be able to tune the parameters of the \emph{models} in such a way that some of the $\boldsymbol{A}^n$ become linearly dependent and thus no longer span the whole phase space. The space of configurations for which the no-slip condition is maintained is then the complement of the space spanned by the $\boldsymbol{A}^n$. The fewer linearly independent $\boldsymbol{A}^n$ there are, the less tuned must be the configuration.

We will limit our discussion to such cases where at least $\ddot{C}$ is linearly dependent on $C,\dot{C}$. The remaining case, where only $\dddot{C}$ is linearly dependent, has only one field-coordinate free, i.e.\ would require initial conditions which are too tuned to be of interest.

The discussion above can be trivially extended to models with more degrees of freedom. Whether the same conclusions would hold, however, is an open question that we will not address here.

\subsection{Horndeski Scalar-Tensor Model\label{sec:Hordenski}}
In this section, we use coordinate time $t$ with its derivative denoted by an overdot and $\ln a$ with the derivative denoted by a prime.

For the Horndeski scalar-tensor/dust system, we use the formulation of the perturbation dynamics described in Ref.~\cite{Bellini:2014fua} where it was shown that linear perturbation theory in an arbitrary Horndeski model can be fully described by specifying a background expansion history $H(t)$, four  functions of time defining the perturbations, $\alpha_\text{K,B,M,T}(t)$ and the initial matter density, which then evolves according to the matter conservation equation. At least one of $\alpha_\text{M,T}$ must be non-zero in order for the propagation of gravitational waves to be modified, see Eq.~\eqref{eq:anisoeq}.  All these functions are in principle arbitrary,
 corresponding to different model choices and background initial conditions, and are complemented by algebraic constraints arising from the requirement of background stability.%
 \footnote{It is possible to construct tuned functions $K,G_i$ to give arbitrary $\alpha_i$, Whether such models are reasonable or typical is a separate discussion, see Ref.~\cite{Linder:2015rcz}.} %
 In particular, we have the requirement that the scalar be dynamical and not be a ghost, 
\begin{equation}
D \equiv \ak + \frac{3}{2}\ab^2 > 0\,,\;\;\ab\neq 2 \,, \label{eq:HornD}
\end{equation}
and that there be no gradient instabilities in the scalar sector (positive sound speed for scalar perturbations)
\begin{align}
	Dc_\text{s}^2 =&(2-\ab)\left(  \frac{\ab}{2}(1+\at)+(\am-\at)  -\frac{\dot{H}}{H^2}\right) - \notag\\ 
	&- \frac{\dot{\alpha}_\text{B}}{H}+\frac{\tilde{\rho}_\text{m}}{H^2} >0 \,. \label{eq:Horncs2} 
\end{align}
Since the effective Planck mass $M_*$ evolves in Horndeski models, it is helpful to define a non-conserved matter density $\tilde{\rho}_\text{m} \equiv \rho_\text{m}/M^2_*$, which evolves according to,
\begin{equation}
\dot{\tilde{\rho}}_\text{m}+3H\tilde{\rho}_\text{m}=-\am H\tilde{\rho}_\text{m} \,. \label{eq:matcons}
\end{equation}

It is most transparent to pick as the basis for the phase space the vector,
\begin{equation}
\boldsymbol{X}= \{\Phi,\dot{\Phi},\Psi, \dot{\Psi}\} \label{eq:HornBasis}\,,
\end{equation}
which is possible provided $\alpha_\text{M}\neq\at$ (we return to exceptions at the end of this section). The no-slip condition \eqref{eq:Aniso0} is then trivial,
\begin{equation}
C = \Psi -\Phi= 0\,. \label{eq:CHorn}
\end{equation}
Since this constraint does not involve any derivatives, $\dot{C}$ is an independent hypersurface, 
\begin{equation}
\dot{C}=\dot\Psi-\dot\Phi = 0 \,,\label{eq:dCHorn}
\end{equation}
and immediately we have that the space of initial conditions satisfying $C=\dot{C}=0$ is at most two dimensional. This case is realized when $\ddot{C}$ is a linear combination of $C$ and $\dot{C}$ on shell, i.e.\ $\ddot{C} \approx \gamma_0 C + \gamma_1 \dot{C}$, where $\gamma_{0,1}$ are some arbitrary functions of time. Let us try to find such a model. 

$\ddot{C}=\ddot{\Phi}-\ddot{\Psi}$ and therefore both the trace Einstein equation and the equation of motion for the scalar must be used to eliminate the second derivatives. On shell, we have
\begin{align}
\ddot{C} - \gamma_1 \dot{C} -\gamma_0 C \approx  A_1 \dot{\Phi} + A_2 \Phi + A_3 \frac{k^2}{a^2} \Phi =0 \,, \label{eq:HornddC}
\end{align}
where the $A_i$ are functions of the background properties and therefore of time, but not of scale.

Rather than using as a basis the four $\alpha_i$ functions, it is helpful to redefine them to the set $(R,S,D,\at)$, with 
\begin{equation}
R \equiv \frac{\am}{\at}\,,\quad S \equiv \frac{\ab}{2} + R - 1  \label{eq:alphabasis}\,.
\end{equation}
and $D$ defined in Eq.~\eqref{eq:HornD}. This is allowed provided $\at\neq 0$. 
The three equations $A_i=0$ can then be written fully as
\begin{align}
S'=&R' -\frac{R}{2}\left(2S(1+\at)+5+3\omp\right)+ \label{eq:gensystem}\\
&+R^2 +\frac{3}{2}(S(1+\omp)+\omm) \,,\notag\\
D'=&\frac{2DR'}{S}+6(R-1)RS-D(R(1+\at)-3(1+\omp))- \notag\\
&-6(1+\at)RS^{2}+\frac{D(3\omm-R(5+3\omp)+2R^{2})}{S}\notag\,,\\
R''=&R'\bigg(\frac{7+9\omp}{2}-(3+\at)R-\notag\\
	&\qquad-\frac{6RS(R-(1+\at)S-1)}{D}\bigg) +\notag\\
&+\frac{3R^{2}S(5-2R+3\omp)(R-1-(1+\at)S)}{D}+\notag\\
&+\frac{1}{2}R\Big((R(1+\at)-1-3\omp)(5-2R+3\omp)\notag\\
&\qquad\quad+3\omp'\Big) \notag\,.
\end{align}
where poles at $D=0$ and $S=0$ motivate the choice of the basis \eqref{eq:alphabasis}. We have redefined the DE pressure $\mathcal{P}$ to the dimensionless variable $\omp\equiv \mathcal{P}/3H^2$. In deriving this system, we have already eliminated the DE energy density using the first Friedmann equation, exchanging it for the matter density fraction $\Omega_{\text{m}}\equiv\tilde{\rho}_\text{m}/3H^2$, while we solve for the derivatives of $H$ using the acceleration equation,
\begin{equation}
2H' +(3+\omp)H = 0 \,. \label{eq:accn}
\end{equation}
The only remaining independent equation is the matter (non-)conservation equation \eqref{eq:matcons}, which  in terms of the matter density fraction is%
\footnote{Remember that $R\at=\am$, so that matter conservation is related to the running of the effective Planck mass.}%
\begin{equation}
	\omm' = \omm \left(3\omp -R\at\right) \,. \label{eq:matterconsomm}
\end{equation}
Finally, choosing $\ln a$ as the time variable (and $'$ to denote the derivative), allowed us to eliminate the explicit dependence on $H$.

Absent any statements regarding the naturalness of models, the dynamics of Horndeski models must satisfy the matter conservation equation for $\omm$, Eq.~\eqref{eq:matterconsomm}, while the only condition on the other variables is that the stability conditions for the scalar and tensor perturbations be satisfied. There always exists some, possibly unnatural, Lagrangian and initial conditions for the scalar field which would give any desired evolution for the other variables, $\omp, \at,D,R,S$. Enforcing dynamical screening in Horndeski requires that the dynamics of the model parameters be restricted to those satisfying three more equations, the system \eqref{eq:gensystem}. We thus have four equations for six variables, and a solution could always in principle be found. Without placing any other requirements (such as no tuning or a choice of a particular background), we can find a Horndeski model which would screen the gravitational slip, while at the same time modifying the propagation of gravitational waves.%
\footnote{More usually, the freedom of Horndeski models is described as \emph{five} functions (background and four $\alpha$ functions) and the matter fraction today $\Omega_\text{m0}$ (e.g.\ \cite{Bellini:2013hea}). This statement makes the implicit assumption of the matter conservation equation \eqref{eq:matterconsomm}, which we have chosen to enumerate here explicitly.}

Nonetheless, the question remains whether these models can in some sense be generic or reasonable. For a tuning that is not too stringent, model parameters should not evolve with timescales too distinct from cosmological, i.e.\ they should be constant or possibly power laws in the scale factor $a$, at least over a short-enough time period such as the matter domination era. 

On the other hand, the equations \eqref{eq:gensystem}, representing trajectories which the model parameters should follow to screen slip, are highly non-linear in the variables $D,R,S$ and therefore the complete phase space contains many timescales for their evolution. So if the evolution of a model were to follow a generic trajectory of the system \eqref{eq:gensystem}, its parameters would have to evolve at a much different rate than $H$. We would then say that this model is very tuned. Equivalently, if we start a slowly-evolving ``natural'' model with a generic initial choice of $D,R,S$, the values of these parameters required to continue to screen slip would rapidly move away from those possessed by the model and the screening would cease. 

Only in the vicinity of the fixed points (FPs) of the phase space of \eqref{eq:gensystem} can the evolution of the trajectories be slow. Exactly at the FPs, the parameters are constant, while close enough to the FPs, the approximate equations are linear and the solutions can be polynomials of $a$. Thus, close to the FPs, for some appropriate linear combination of the model parameters $v$, the screening trajectory requires,
\begin{equation}
v(a) = v_* + v_1 a^n \,,
\end{equation}
which is slow enough by our definition and an appropriate untuned model can be found. Ideally, we would find attractor FPs, with $n<0$, since this would allow for an indefinite duration of screening for this choice of model parameters. However, even unstable FPs allow for a slow evolution of the model parameters for some short enough period of time, with the duration dependent on how closely the initial conditions would lie to the FP.  Our aim is to find and categorize these fixed points, identifying the reasonable model parameters required to screen slip. 

The system \eqref{eq:gensystem} contains two external parameters which are arbitrary: $\omp$ and $\at$. We wish to concentrate on the time scales internal to the $D,R,S$ system \eqref{eq:gensystem}, so we choose to remove those to do with the external parameters. For the analysis, we assume a constant $\at$ and a pressure which is such that the matter fraction is conserved, compensating for the evolving Planck mass with a modification of the background, $\omp=R\at/3$. The latter choice makes $\omm$ a constant external parameter. On the other hand, this particular choice for $\omp$ does mean that the expansion history \eqref{eq:accn} does not precisely correspond to matter domination. Choosing a small enough $\at$ would bring us back close enough to satisfy any observable constraints.%
\footnote{Note that it will turn out that $R=0$ will be the coordinate of one of the fixed points, FP1, and therefore the exact matter-domination case is included in our analysis.} %
The coordinates and nature of the fixed points we find will be functions of the external parameters $\at$ and $\omm$. We could then reintroduce a slow evolution of these parameters and would find that the fixed points would move adiabatically without changing their nature, until an FP stability criterion were violated. In fact, we choose to present the case $\omm=1$, since changing this value does not introduce any qualitatively new information.

Thus the system we actually study is the three equations \eqref{eq:gensystem} with the constant external parameter $\at$, $\omm=1$ and the pressure $\omp=3R\at$. We will refer to it as the \emph{MD system}.
%

The MD system has three fixed points with coordinates $(D,S,R,R')$:
\begin{align}
\text{FP1: }&(D,-1,0,0)\,,\label{eq:HornMDFPs}\\
\text{FP2: }&\left(-\frac{9(7-\alpha_{T})\alpha_{T}(4+3\alpha_{T})}{8(1+\alpha_{T})(1+2\alpha_{T})^{2}},
\frac{3(2-\alpha_{T})}{4(1+2\alpha_{T})},
\frac{5}{2-\alpha_{T}},0\right)\,,	\notag\\
\text{FP3: }&\left(\frac{9(2+\alpha_{T}(4+3\alpha_{T}))}{2(1+2\alpha_{T})^{2}},-\frac{1}{\alpha_{T}}-\frac{3\alpha_{T}}{2+4\alpha_{T}},-\frac{1}{\alpha_{T}},0\right)\,. \notag
\end{align}
FP1 is in fact a line independent of the value of $D$ and is also the only fixed point of a system with true matter domination, $\omp=0$;  the sound speed is positive for subluminal GWs, $-1<\at<1$. Allowing for an evolving Planck mass with a scaling solution has added FP2 and FP3. However, FP3 is always either a ghost or has negative sound speed, while FP2 is healthy only in the range $7.0<\at<15.0$, which is outside of that implied by the LIGO measurements \cite{Blas:2016qmn}. Allowing for $\omm<1$ increases the upper bound for stability at FP1 to $1-\omm$, so that GWs are allowed to be superluminal when a non-vanishing density fraction of DE is present. It also increases the allowed range of $\at$ for FP2.

Linearizing the system around the fixed points shows that none of them are fully stable: for $\at>-1$, FP1 has no stable directions, FP2 has two unstable directions for $\at<2$ and 3 otherwise, and FP3 --- one for $-\frac{1}{2}<\at<0$ and two otherwise. 

In summary, only FP1 is of potential interest, representing a model with $\am=\ab=0$, $\ak>0$ and a modification of tensor speed. It is however an unstable FP, thus it will offer an approximate solution only for a limited time. Given sufficient tuning of the model's action so that the initial model parameters are sufficiently close to FP1, this period could of course be made as long as is required, to cover all of matter domination and possibly could be connect to an acceleration era, but we set out to avoid such a tuning and will not consider it further. 
\\

A general trajectory will rapidly move away from any of these fixed points toward infinity. As we detail below, we have studied the full non-linear dynamics numerically for some choices of parameters, confirming that none of these fixed points are end points of evolution for the required model-space trajectories. We find that for $-\frac{1}{2}<\at<2$, essentially all the trajectories flow toward $D\rightarrow \pm \infty$ while $S$ and $R$ converge to fixed values given by
\begin{equation}
		S_* \equiv \frac{3(2-\alpha_{T})}{4(1+2\alpha_{T})}\,,\quad R_* \equiv \frac{5}{2-\alpha_{T}} \label{eq:limits}\,,
\end{equation}
In the limit $|D|\rightarrow \infty$, when linearized around $S=S_*$, $R=R_*$, $R'=0$, the MD system reduces to
\begin{align}
D'&=\frac{5(1+\alpha_{T})}{2-\alpha_{T}}D \label{eq:HornLinSys}\\
\delta S'&=-\frac{2(1+2\alpha_{T})}{2-\alpha_{T}}\delta S+\frac{8+\alpha_{T}(40+3\alpha_{T})}{8(1+2\alpha_{T})}\delta R+\delta R'  \notag\\
\delta R''&=-\frac{\left(16-3\alpha_{T}\right)}{2(2-\alpha_{T})}\delta R'+\frac{5+3\alpha_{T}}{2(2-\alpha_{T})}\delta R \notag
\end{align}
where $\delta S\equiv S-S_*$ and $\delta R\equiv R-R_*$. In the $(S,R,R')$ subspace, the linearized system has eigenvalues
\begin{equation}
		4-\frac{10}{2-\alpha_{T}},-\frac{5}{2},1-\frac{5}{2-\alpha_{T}}
\end{equation}
which are all negative if $-\frac{1}{2}<\at<2$. This confirms the behavior we have observed in the numerical study.

At $S_*,R_*$ and $|D|\rightarrow\infty$, the product $Dc_\text{s}^2>0$ provided that $\at>0$. Thus on trajectories with $D>0$ (which is possible, depending on initial conditions), the model provides a background which is stable to perturbations when gravitational waves are superluminal. Note, however, that this asymptotic configuration does have not a precise matter dominated background, since $\omp=R_*\at/3$. As $R_*$ is finite, the background could nonetheless be chosen to be close enough to observational constraints by picking a sufficiently small $\at$.

Thus we have in principle found a limiting ``attractor'' Horndeski model parameters which would have to be approached during matter domination if the no-slip condition were to be preserved at all scales:
\begin{align}
	&\ak \rightarrow \infty \,, 	&&	0<\at<2 \,, \\
	&\ab = 2-\frac{10}{2-\at} + \frac{3(2-\at)}{2(1+2\at)} \,,  &&\am = \frac{5\at}{2-\at} \notag\,.
\end{align}
Moreover, when this limiting model is taken, the evolution equation for the gravitational slip becomes
\begin{equation}
	\ddot{\Sigma} + \frac{2(4+\at)}{2-\at}H\dot\Sigma+\frac{5(3+4\at+\at^2)}{(2-\at)^2}H^2 \Sigma=0 \,, \label{eq:HornPiEvol}
\end{equation}
with $\Sigma\equiv \Psi-\Phi$, while $\Psi\rightarrow \text{const}$ and therefore, for $0<\at<2$, it does decay away over time irrespective of the initial conditions. We can thus satisfy both our requirements for dynamical shielding during matter domination. 

Given this understanding, we could now take step back and take the $D\rightarrow \infty$ limit of the original system \eqref{eq:gensystem} and modify the pressure to include a cosmological constant-like component, $\omp=R\at/3 - (1-\omm)$. When this is done, trajectories evolve from the matter-domination FP \label{eq:limits} to a future acceleration attractor with some new but finite values of $R$ and $S$.

However, we need to stress that $D\rightarrow \infty$ is a very peculiar limit, in which the kinetic energy of the fluctuations becomes infinite and therefore the scalar field essentially ceases to be dynamical. If the action is considered for a rescaled field absorbing this divergence, $\pi\equiv \sqrt{D}\delta\phi/\dot{\phi}$, one can see that the sound speed vanishes and all the interactions vanish. This is a limit in which the quantum fluctuations diverge (since they are normalized as $1/\sqrt{c_\text{s}}$) and thus the classical theory should not be taken at face value. Our opinion is that this is a very pathological situation and it is not worth studying further, despite the fact that the naive linear perturbation theory behaves in the way we asked.


\subsubsection*{Numerical analysis}

To understand the non-linear dynamics of the system, we have carried out a numerical study of the MD system of equations. Although a fully analytical statement is always desirable, our numerical study confirms the dynamics as we have described them above.
 For the initial conditions (at $t =t_{\text{min}}$) we constructed a grid spanned by $D= \mathcal{I}_{+}$, 
$S=\mathcal{I}_{+} \cap \mathcal{I}_{-}$, 
$R=\mathcal{I}_{+} \cap \mathcal{I}_{-}$, with $\mathcal{I}_{-}  = \{-10^{-10}, -10^{-10 + \delta n}, \ldots, -10^{n_{\text{max}}} \}$, $\mathcal{I}_{+} = \{10^{-10} ,  10^{-10 + \delta n}, \ldots, 10^{n_{\text{max}}} \}$ with the exponent's step set to $\delta n = 0.6$ and $n_{\text{max}} = 1$. $R'$ was fixed at the value $R'=10^{-5}$. The numerical simulation was carried for three values of $\at$:
\be
\at = \{-0.7, 0.6, 3 \},
\ee 
while the integration time was set between $t_{\text{min}} = \text{log} (10^{-25})$ and $t_{\text{max}} = 0$. For each value of $\at$, we calculated $\mathcal{N} = 25992$ trajectories by employing an implicit Euler method with a fixed step of $1/500$.

We focus on the case $\at  = 0.6$,  it since lies in the range for which the fixed points \eqref{eq:HornMDFPs} have the most attractive directions. In $98.9\%$ of trajectories calculated, $D$ diverged (with half towards $+\infty$), while $R \rightarrow 3.57$ ($R' \to 0$) and $S \rightarrow 0.48$ as $t \to t_\text{max}$. These are the limits predicted in Eqs~\eqref{eq:limits}. The remaining $1.1\%$ of trajectories evolved toward $D\to 0$, but a closer inspection of the evolution shows that the numerics suffer from instability in the vicinity of the $D=0$ pole of the MD system and we conclude that these are not proper solutions.

For the other two values of $\at$, lying either side of the interval with maximum stable directions around fixed points, we find that $99\%$ of trajectories evolve toward a diverging $D$ and $S$, with $R\rightarrow R_*$ as given by results~\eqref{eq:limits}. The remaining $1\%$ of trajectories appear to evolve toward $D=0$, but again these solutions are marred by numerical instability and we do not consider them trustworthy.


\vspace{0.5cm}

As we pointed out earlier, the choice of basis for the fields \eqref{eq:HornBasis} and model parameters \eqref{eq:alphabasis} is inappropriate for certain $\alpha_i$. We have investigated these cases separately and find that they do not provide viable models:
\begin{itemize}
	\item $\am=\at$: For this choice, the no-slip configuration $C=0$ implies that $\Phi=0$ and thus the gravitational slip is shielded only when there is no gravitational field at all, making this case uninteresting.
	
	\item $\at=0$: Preserving the no-slip configuration under time evolution requires that $\ab=2\am$. We then have $D=\ak + 6\am^2$ and $Dc_\text{s}^2 = -\frac{D}{3}$. It is thus impossible to choose the remaining independent $\alpha_i$ to give a model which is simultaneously not a ghost and has a positive sound speed squared.	
\end{itemize}

\vspace{0.5	cm}

To summarize, we have shown that scalar-tensor theories are in principle flexible enough to allow for models which achieve dynamical shielding of gravitational slip. For generic cases, the model parameters need to evolve on timescales much faster than cosmological which signifies a very special solution. Then, there exists one class of models which can screen for a limited time, but the parameter values need to be tuned very precisely to make this long enough for a realistic cosmology. The only model class which naturally screens for an extended time requires a divergent kinetic term for the scalar field and therefore vanishing sound speed. This transforms the scalar into a kind of dust. This should have been expected of course, since we are essentially requiring that the scalar field follow the evolution of dark-matter dust in a very precise manner.
\subsubsection*{Quasi-Static Horndeski}
Frequently, the quasi-static (QS) approximation is used to model the evolution in scalar-tensor modifications of gravity. One neglects the homogeneous part of the solution for the scalar field, i.e.\ one assumes that the scalar is not dynamical but rather follows the matter distribution. This can be justified when the scales under consideration are both inside the sound horizon of the scalar and the cosmological horizon \cite{Sawicki:2015zya}. 

In Horndeski models, $C=0$ does not contain time derivatives or different weights in $k^2$ and thus is unaffected by the QS approximation. The QS approximation in our case is equivalent to enforcing only $A_3=0$  in Eq.~\eqref{eq:HornddC} on the model space while neglecting the tuning required by $A_1=A_2=0$, i.e.\ we require
\begin{equation}
	\at = \frac{H^2\am(\ab+2\am)}{(2-\ab)H^2\am+\dot{H}(2-\ab)-H\dot{\alpha}_\text{B}+\tilde{\rho}_\text{m}} \,.
\end{equation}
This is the same as the QS linear shielding conditions proposed in Ref.~\cite{Lombriser:2014ira}, although there the condition $A_2=0$ is also enforced to cancel the subdominant sources outside the sound horizon. Provided that the sound speed of the scalar is close to that of light, this QS solution is approximately valid inside the cosmological horizon. However, in the vicinity of the sound horizon, the QS approximation breaks down \cite{Sawicki:2015zya} and the deviations from the no-slip configuration become large. Indeed when a mode cross the sound horizon, it takes time before it decays to its QS approximation. Thus even inside the horizon, the slip would vanish only up to corrections of order  $\text{max} \left(aH/c_\text{s}k,aH/k\right)$. 

In particular, in Ref.~\cite{Linder:2014fna}, the covariant Galileon model was studied in the QS limit, demonstrating that for appropriate choices of parameters, the slip does vanish at particular moments in time. This vanishing is exactly a QS screening of the sort described above and not a counter-example of the arguments presented in this paper: there would be suppressed corrections to the slip everywhere and it would reappear at scales close the horizon. Nonetheless, given the power of any conceivable instruments at large scales and cosmic variance, this may well be never detectable for many models. On the other hand, in Ref.~\cite{Lombriser:2015sxa}, an even stronger QS shielding condition is chosen, where also the effective Newton's constant for perturbations takes the standard value and the authors find that when the full dynamics is solved the oscillations in the scalar would be unobservably small.

\subsection{Einstein-Aether theories\label{sec:EA}}

Now we ask the same question for the case of EA theories: Does a modified propagation of gravitational waves always imply the existence of a gravitational slip? As we show, within EA theories, there exists a one-parameter family of models which preserves a no-slip configuration if one is set up initially. However, the gravitational slip is conserved in these models, so if the initial configuration has it, it will remain.  Thus, EA models cannot provide a dynamical shielding mechanism sufficient for our requirements.

Let us start by briefly reviewing the dynamics of the theory at the linear level of scalar perturbations around FLRW, following the equations of Ref. \cite{Lim:2004js}.\footnote{Notice the difference in our convention for the gravitational potentials $\Phi$ and $\Psi$ compared to \cite{Lim:2004js}.} 

In the presence of dust, the model describes two propagating scalar degrees of freedom: one for the dust, the other -- for the helicity-zero component of the vector. The conservation equation for the energy-momentum tensor for the vector yields two equations: a constraint equation for the lagrange multiplier $\lambda$, and a dynamical equation for the vector respectively. The former allows one to eliminate $\lambda$ from the rest of the equations, and using Einstein equations one can arrive to a set of two second-order equations for two dynamical fields, which can be conveniently chosen to be the spatial divergence of the vector perturbation $\Theta$, defined in Eq.~\eqref{eq:ThetaDef} and the Newtonian potential $\Phi$. The equation of motion for $\Theta$ takes the following schematic form
\begin{align}
& \Theta'' +  \mathcal{E}_{\Theta'}(\beta_i)\mathcal{H} \Theta' + \left(\mathcal{E}_{\Theta (1)}(\tau; \beta_i) + \mathcal{E}_{\Theta (2)}(\beta_i)\frac{k^2}{\mathcal{H}^2} \right) \mathcal{H}^2 \Theta  \notag \\
&=  \frac{k^2}{\mathcal{H}^2} \left( \frac{2 \beta_1 + 3\beta_2  + \beta_3}{\beta_1 \omega}\right) \left( \mathcal{H}^3 \Phi + \mathcal{H}^2 \Phi'   \right) ,\nonumber \\
 \label{Theta''}
\end{align}
with the explicit form of the dimensionless coefficients $\mathcal{E}_{i}$ to be read from Ref.~\cite{Lim:2004js} and $\omega \equiv ( 1 - 16 \pi Gm^2 (\beta_{1}+\beta_{3}) )$. A similar equation follows for $\Phi$ from the trace of the $ij$ part of the Einstein equations. Note that the perturbation of the dust density does not appear in the equation of motion for $\Theta$ if one assumes that the corresponding energy-momentum tensors are separately conserved. The configuration of this theory is thus specified by the set 
\begin{equation}
\boldsymbol{X}= \{\Theta, \Theta', \Phi, \Phi' \}. \label{eq:EAvars}
\end{equation}

The anisotropy constraint is 
\begin{align}
k^2(\Psi-\Phi) = 16 \pi G m^2 (\beta_1 + \beta_3) \left(2  \mathcal{H} \Theta + \Theta' \right). \label{EA:AnisoEq}
\end{align}
 and thus the requirement of zero anisotropic stress in this case is equivalent to requiring at least one of the following:
\begin{align}
&i) \, m^2 = 0, \; \; \;  ii) \,  \beta_1 + \beta_3 = 0, \; \; \; iii) \, (a^2 \Theta)^{'} = 0.
\end{align}
The first two are conditions on the theory space of the model, while the last one places a constraint on the linear variable $\Theta$. In particular, the first condition essentially restores the Lorentz symmetry of the theory (setting the v.e.v of $u^\mu$ to zero) and if imposed, all the non-trivial contributions from the vector vanish in the linearized equations, reverting the theory to GR. The second choice will eliminate the shear term in the action and therefore any modification to GW propagation. The third condition is what  interests us, since it is a condition on the configuration variables \label{eq:EAvars}, which, if preserved, would imply that gravitational slip vanishes. Let us investigate it further below. 

We want to show whether the no-slip condition
\be
C[\boldsymbol{X}] \equiv (a^2 \Theta)' = 0, \label{EA:C=0}
\ee
is preserved under evolution in time, i.e.\ whether 
\be
C' + \gamma_0 C \approx 0, \label{EA:C'=02}
\ee
with $\gamma_0$ some function of time and the $\beta_i$.

Condition \eqref{EA:C'=02} defines a second-order differential equation in time for $\Theta$, which has to be satisfied on-shell. We use the equation of motion (\ref{Theta''}) in (\ref{EA:C'=02}) to eliminate $\Theta''$. This leads to a new constraint between $\Theta$, $\Phi$ and their first time derivatives, which has to be satisfied at all times and scales in the cosmological evolution:
\begin{align}
C' + \gamma_0 C &\approx  
 \frac{a^2}{\beta_{1} \omega}  \Bigg[ \left(\frac{ \mathcal{H} '}{ \mathcal{H}^2} - 1 \right) \left(2\beta_2 + 3\beta_2 + \beta_3 \right)  \label{EA:C'=03} \\
 & - \frac{k^2}{ \mathcal{H}^2}(\beta_1 + \beta_2 + \beta_3) \Bigg]  \mathcal{H}^2 \Theta + \notag \\
 & + \frac{a^2 k^2}{ \mathcal{H}^2} \frac{\left( 2\beta_2 + 3\beta_2 + \beta_3\right)}{\beta_1 \omega} \left(   \mathcal{H}^3 \Phi +  \mathcal{H}^2 \Phi' \right) \notag
 \,.
\end{align}
This constraint can be satisfied for all no-slip configurations provided that 
\begin{align}
& \beta_1 + \beta_2 + \beta_3 = 0 \label{ConditionEAbeta1} \,,\\
& 2\beta_1 + 3\beta_2 + \beta_3 = 0 \label{ConditionEAbeta2} \,.
\end{align}
Conditions (\ref{ConditionEAbeta1}) and  (\ref{ConditionEAbeta2}) are satisfied simultaneously when
\begin{align}
\beta_{1} = -2 \beta_{2}, \; \; \; \beta_{3} = \beta_{2}. \label{ConditionEAbeta3}
\end{align}
The choice (\ref{ConditionEAbeta3}) defines a family of models parameterized by $\beta_2$, for which the no-slip condition is preserved under time evolution. This family of models then has a vanishing sound speed for the scalar and a modified GW speed:
\begin{align}
c_{\text{s}}^2 &= \frac{\beta_{1} + \beta_2 + \beta_3}{\beta_1 + 16 \pi Gm^2 (\beta_{1} + \beta_{3})(\beta_{1}+3\beta_{2} + \beta_{3})} = 0 \,,
\\
c_{\text{T}}^2 &= (1-\beta_1-\beta_3)^{-1} = (1+\beta_2)^{-1} \,. \notag
\end{align}

We have thus shown that, in  models (\ref{ConditionEAbeta3}), configurations with vanishing gravitational slip are maintained under time evolution. The second question is whether such no-slip conditions are typical in these models. In models (\ref{ConditionEAbeta3}), the equation of motion (\ref{Theta''}) reduces to
\be
C' = 2\mathcal{H} C \,,
\ee
and therefore that $C\propto a^{2}$. Eq.~\eqref{EA:AnisoEq} then implies that
\begin{equation}
\Psi-\Phi = \text{const}	\,.
\end{equation}
Contrary to the Horndeski case Eq.~\eqref{eq:HornPiEvol}, in this EA model, if there were any slip in the initial configuration, it would be conserved for all time. Thus EA models do not generically evolve to a no-slip configuration and will generically exhibit gravitational slip. 

Moreover, just as is the case with such Horndeski theories, the models \eqref{ConditionEAbeta3} have zero sound speed. This leads to problems with normalizing quantum fluctuations, which is exacerbated here since the parameters $\beta_i$ are constant, implying that the sound speed for the helicity-zero mode was zero also during inflation.

We thus conclude that there are no Einstein-Aether models in which the dynamical shielding mechanism would generically produce a universe with no gravitational slip but with modified GW speed.

\subsection{Bimetric theories} \label{sec:Bimetric}
\subsubsection{Background}

We now study the massive bigravity theories with matter minimally
coupled to the metric $g$. We assume both metrics to be spatially flat 
FLRW at the background level
\begin{eqnarray}
\bar{g}_{\mu\nu}dx^{\mu}dx^{\nu} & = & a^{2}(-d\tau^{2}+\delta_{ij}dx^{i}dx^{j}),\\
\bar{f}_{\mu\nu}dx^{\mu}dx^{\nu} & = & b^{2}(-c^{2}d\tau^{2}+\delta_{ij}dx^{i}dx^{j}),
\end{eqnarray}
where $a$ and $b$ are the two scale factors. We also define their ratio $r\equiv b/a$. Since the two metrics have in principle independent time coordinates, the lapse $c(\tau)$ remains after choosing conformal time $\tau$ of the metric $g$ as the time coordinate. We also have two a priori independent conformal Hubble parameters, $\cH \equiv a'/a$ and $\cH_f\equiv b'/cb$.

In this model, a part of the diffeomorphism group is broken, so matter conservation must be imposed externally. In such a case, satisfying the background Bianchi identities leads to two branches of
solutions: (i) the \emph{algebraic} branch where $r=\text{const}$ and the dynamics on FLRW are strongly coupled (helicity-0 and helicity-1 of the massive graviton do not propagate); and (ii) the \emph{dynamical} branch, where the Bianchi identities constrain the lapse $c$ to obey the relation 
\begin{equation}\label{bianchi}
c=1+ \frac{r'}{\cH r}\,,
\end{equation}
which implies $\mathcal{H}_f=\mathcal{H}$, see Ref.~\cite{Comelli:2012db,Konnig:2013gxa} for details. Since it is the only possibly healthy configuration, we only consider the \emph{dynamical} branch henceforth. 

We can define four polynomials of $r$ with coefficients linear in the constants $\beta_n$ introduced in the action \eqref{eq:bigravAction},
\begin{align}
\rho_{g} & \equiv  \beta_{0}+3\beta_{1}r+3\beta_{2}r^{2}+\beta_{3}r^{3}\,,\\
\rho_{f} & \equiv  \beta_{4}+3\beta_{3}r^{-1}+3\beta_{2}r^{-2}+\beta_{1}r^{-3}\,,\\
Z & \equiv \beta_{1}+\beta_{3}r^{2}+2\beta_{2}r\,,\label{eq:defZ}\\
\bar{Z} & \equiv  \beta_{1}+\beta_{3}cr^{2}+\beta_{2}(c+1)r\,, \label{eq:z-bar}
\end{align}
The functions $\rho_{g}$ and $\rho_{f}$ play the role of the background energy densities arising from the interaction potential \eqref{eq:bigravAction} in the $g$ and $f$ Friedmann equations. The pressure from this potential depends in addition on the combination $Z$ as well as the lapse $c$. The two pairs of Friedmann equations can be combined to find the expressions
\begin{eqnarray}
\cH' & = & \frac{a^{2}c\rho_\text{m}}{2(c+2)}-\frac{a^{2}(c-1)Z\left(cr^{2}+2\right)}{2(c+2)r},\label{eq:bi_hubbledot}\\
\mathcal{H}^{2} & = & \frac{a^{2}r^{2}\rho_{f}}{3}\label{eq:bi_hubble}\\
\rho_\text{m} & = & \rho_{f}r^{2}-\rho_{g},\label{eq:bi_rho}\\
c & = & \frac{3\left(1-r^{2}\right)Z+r\left(r^{2}\rho_{f}+3\rho_{g}\right)}{3\left(1-r^{2}\right)Z-2r^{3}\rho_{f}}\label{eq:bi_lapse}\,,
\end{eqnarray} 
where we have restricted the matter content to be dust. Thus all the \emph{instantaneous} background dynamics can be recast in terms of the variables $\{a,r,\rho_{g},\rho_{f},Z\}$. Moreover, the evolution of $a$ is given by Eq.~\eqref{eq:bi_hubble}, $r$ by Eq.~\eqref{bianchi}, $\rho_{f,g}$ by their conservation equations. Thus given that
\begin{equation}
Z' = 2\cH(\bar{Z}-Z)\,, \label{eq:z-bar2}
\end{equation}
and that $\bar{Z}'$ can be similarly recast, the model basis set $\mathcal{B}=\{a,r,\rho_{g},\rho_{f},Z,\bar{Z}\}$ is closed under time evolution and completely specifies the evolution of the background of the model. In fact, no background functions beyond $\mathcal{B}$ enter the coefficients in linear perturbation equations.

We should note that we have simply traded the five constants $\beta_n$  for five other variables with a fixed time evolution. The advantage of this new formulation is that the expressions are more compact and are explicitly related to physical parameters such as the graviton mass $\mu^2=r\bar{Z}$.%
\footnote{The massive and massless fluctuations at linear level correspond to linear combinations of $h_{ij}$ and $\gamma_{ij}$~\cite{Hassan:2011zd}.} %
A non-vanishing pressure of the matter $p_\text{m}$ would provide an additional external variable which needs to be specified, but does not change the discussion otherwise. In what follows we will in fact continue to use expressions contaning the extended set $\left\{ a,c,r,\rho_\text{m},{\cal H},\bar{Z},Z\right\}$, since this is somewhat simpler, with the understanding that $c,\mathcal{H}, \rho_\text{m}$ are really given by Eqs~(\ref{eq:bi_hubble}--\ref{eq:bi_lapse}).

We also point out that in Refs~\cite{Konnig:2014xva,Cusin:2014psa} it was shown that $r'=0$, $c=1$ in the dynamical branch corresponds to the future de Sitter attractor and therefore will not be considered as a relevant background for our proof.

\subsubsection{Perturbations}

Small fluctuations propagate on the background metrics, $g_{\mu\nu}  =   \bar{g}_{\mu\nu}+h_{\mu\nu}$,
$f_{\mu\nu}  =   \bar{f}_{\mu\nu}+\gamma_{\mu\nu}$, 
the scalar part of which we can described using eight scalars when the gauge is not fixed,
\begin{equation}
\begin{array}{ccl}
h_{00} & = & -2a^{2}\Psi_{g},\\
h_{0i} & = & -a^{2}\partial_{i}B_{g},\\
h_{ij} & = & 2a^{2}\left(-\Phi_{g}\delta_{ij}+\partial_{i}\partial_{j}E_{g}\right),
\end{array}
\end{equation}
\begin{equation}
\begin{array}{ccl}
\gamma_{00} & = & -2c^{2}b^{2}\Psi_{f},\\
\gamma_{0i} & = & -cb^{2}\partial_{i}B_{f},\\
\gamma_{ij} & = &2 b^{2}\left(-\Phi_{f}\delta_{ij}+\partial_{i}\partial_{j}E_{f}\right).
\end{array}
\end{equation}

As was shown in \cite{Comelli:2012db}, the gravitational slip in the matter $g$ metric is
\begin{equation}
C = a^{2}  r \bar{Z} \Delta E.
\end{equation}
where $\Delta E \equiv E_f-E_g$ is a gauge-invariant variable. In order to fulfill the no-slip condition $C =0 $ one of the two following conditions must be satisfied
\begin{eqnarray}
& (i)\quad &  r \bar{Z} = 0,\\
& (ii)\quad &\Delta E = 0.
\end{eqnarray}
and preserved on equations of motion.

\paragraph*{Case (i).} Since according to Eq.~\eqref{sec:Bimetric}, $r\bar{Z}$ is the graviton mass, in this case the propagation of gravitational waves is not modified, confirming our hypothesis. Nonetheless, it is interesting to ask when the mass of the graviton can vanish in cosmology.

Since $r\neq 0$, we use Eq.~\eqref{eq:z-bar} to solve for $\bar{Z}=0$ obtaining
\begin{equation}
c=-\frac{\left(\beta_{1}+\beta_{2}r\right)}{\left(\beta_{2}+\beta_{3}r\right)r}\,.
\end{equation}
Combining this with the expression for $c$, Eq.~\eqref{eq:bi_lapse}, gives a polynomial 
in $r$ and $\beta_n$, implying that $r$ must be a constant. The Bianchi constraint \eqref{bianchi} then implies that $c=1$ and the Universe is de Sitter. Thus, in bigravity models on cosmological solutions, the graviton can only be tuned to be massless on exact de Sitter.\\
\\

\paragraph*{Case (ii).}

The second case is a condition on the configuration of the perturbation variables. Requiring that it be preserved under time evolution places a constraint on the model parameters.

The full perturbation equations were first studied in Ref.~\cite{Comelli:2012db} in the gauge-invariant formalism. For our purpose, however, the formulation of Ref.~\cite{Lagos:2014lca} is simpler.  There it was explicitly shown that fixing the gauge such that the matter fluctuations and $\Phi_f$ are set to zero allows one to eliminate all the auxiliary variables, obtaining the full dynamics in terms of just the fields $E_{g}$ and $E_{f}$. The full equations are presented in Ref.~\cite{Lagos:2014lca} and we will not quote them here.  Thus the set of independent linear variables which fully characterizes scalar perturbations in this theory is therefore 
\begin{equation}
\boldsymbol{X} = \{E_g,E_g',E_f,E_f'\}\,.
\end{equation}
The evolution of the full matter-bigravity scalar sector is described by two second-order differential equations, confirming explicitly that in the dynamical branch there are two degrees of freedom: one from the matter, the other -- the propagating helicity-0 mode of the massive graviton.

Since $\Delta E\equiv E_{f}-E_{g}$ is gauge invariant, this expression is also valid in the gauge choice of Ref.~\cite{Lagos:2014lca}. Thus requiring there be no gravitational slip, $C=0$, in this gauge is simply the requirement that $E_g=E_f$. The slip is not generated when 
\begin{equation}
C'=a^2 r\bar{Z} (E_f'-E_g') + \gamma_0 C = 0
\end{equation}
with $\gamma_0$ some function of time. This is a new condition, since $r\bar{Z}\neq0$. Thus the best-case scenario has $C''$ on shell be a linear combination of the constraints $C,C'$

Taking the $\Delta E=\Delta E'=0$ initial configuration, it is enough for us to require that $\Delta E''=0$ be identically satisfied on equations of motion. This is only possible if the equations of motion for $E_{g,f}$ are identical for these initial conditions, i.e.

\begin{align}
E''_{g}  +& \cH E'_{g}-\frac{1}{2}a^{2}\rho_\text{m} E_{g}=0 ,\label{eq:large-k-g-2}\\
E''_{f}+& \cH E'_{g}+ k^{2} A_{k} E_{g}-\frac{1}{2}a^{2}A_{l}E_{g}=0,\, \label{eq:large-k-f-2}
\end{align}
where it will be enough to consider just the small-scale limit, $k\rightarrow\infty$, and where
\begin{eqnarray}
A_{k} & = & \!\frac{(c-1)}{3Z}\left(2\bar{Z}-Z\right),\\
A_{l} & = & \!\frac{(2\bar{Z}\!-\!Z)\rho_\text{m} r+Z(1-c)\!(r^{2}\!+1)\!(2\bar{Z}\!-\!Z(c+1)\!)}{rZ}.\hspace{0.9cm}
\end{eqnarray}
Eq.~\eqref{eq:large-k-g-2} does not contain a $k^2$ term reflecting the fact that we are using dust for the matter, the sound speed of which is zero. As we will see, already in this limit the conditions we require cannot be satisfied and therefore we do not need to consider the subleading corrections.

Equations \eqref{eq:large-k-g-2} and \eqref{eq:large-k-f-2} are identical when 
\begin{alignat}{2}
(I)\quad & A_{k} & = & \:0,\\
(I\!I)\quad & A_{l} & = & \rho_\text{m}.
\end{alignat}
Thus $(I)$ is satisfied for $c=1$, which is an empty de Sitter and not relevant, 
or $2\bar{Z}=Z$. Taking the latter solution, $(I\!I)$ gives
\begin{eqnarray}
\frac{(c-1)\left(r^{2}+1\right)c Z}{r} & = & \rho_\text{m}.\label{eq:biconstraint}
\end{eqnarray}
Using Eqs~\eqref{eq:bi_rho} and \eqref{eq:bi_lapse}, we see this constraint is again just an algebraic equation for $r$, meaning that $r$ is constant. This again implies that $c=1$ and therefore can only be satisfied on de Sitter. 

We thus have proven that if matter is coupled to one of the metrics, no model of bimetric gravity in which the gravitational slip is shielded and yet the graviton has a mass for gravitational waves exists.

\section{Conclusions and Implications}\label{sec:Conclusions}

In this article, we revisited the question of the relationship between the propagation of the gravitational waves and presence of gravitational slip in the scalar sector in theories of modified gravity. In three broad classes of theories, Horndeski scalar-tensor, Einstein-Aether and bimetric gravity, the appearance of gravitational slip from perfect-fluid matter is only possible if the propagation of gravitational waves is modified. On the other hand, beyond-Horndeski models are the first example of models where this relationship is violated: gravitational slip can be sourced without any corrections to gravitational waves. In this respect, this modification of gravity behaves more like a correction to the matter energy-momentum tensor.

The majority of this paper, however, was devoted to investigating to what extent it is possible to pick \emph{model} parameters so that GW propagation be modified implying that there could be gravitational slip, but it nonetheless be dynamically shielded at all scales. We have shown that both the bimetric and Einstein-Aether models are too rigid for such a requirement. However, the scalar-tensor models can be tuned to permit such shielding. 

In the case of Einstein-Aether, there exists a family of models, parameterized by the speed of GW, where a no-slip initial configuration is maintained under evolution in time. However, if there is any slip in the initial configuration, it is preserved. Moreover, the sound speed of the helicity-zero mode is exactly zero, so the initial conditions that one might set during inflation is actually completely out of control. 

Horndeski models are much more flexible, since the model parameters are in principle allowed to be functions of time. Indeed, we have shown that for any chosen cosmological background, it is possible to tune the evolution history of the parameters in such a way that a no-slip condition be maintained under evolution in time. However, shielding with generic choices of model parameters requires that their evolution be much faster than cosmological, signifying the existence of fine tuning or cancellations in the action. As such, this general solution does not point toward a physically interesting model. We attempted to find a more generic model, one which did not have to be precisely tuned to maintain screening for a long time. We found that such a setup requires a divergent kinetic term for the scalar, implying that one cannot write down a proper action within the Horndeski class for such a model.

Moreover, this setup again produced a model which has a zero propagation speed for the scalar mode, and therefore which suffers from the same issues as the Einstein-Aether case. This is not altogether surprising, since we are requiring that the scalar perturbations mimic the dynamics of the dust exactly, following the collapse of structures in the universe. Notwithstanding all these pathologies, in this model, any gravitational slip present in the initial conditions during matter domination does actually decay away, so the dynamics would shield the gravitational slip as we have required.

If only an approximate statement is required, it is possible to choose the Horndeski model parameters in such a way that,  in the quasi-static regime, this shielding of slip is effective. The tuning required for this is much less onerous. However, this approximation relies on neglecting time derivatives. Taking this approach, one misses out the fact that this tuning nonetheless cannot suppress corrections to the gravitational slip of order $\text{min}(\mathcal{H}/k,\mathcal{H}/c_\text{s}k)$ which become very significant at scales close to the cosmological/sound horizon. Thus even if the model of gravity happens to have such parameters that the gravitational slip is suppressed at smaller scales, its measurement near horizon scales would still be informative. The question of course is to what precision Euclid or the Square-Kilometer Array will be capable of measuring this property, especially if it depends on scale \cite{Amendola:2013qna,Alonso:2015sfa,Pogosian:2016pwr}.

Indeed, the cosmic microwave background is very much a near-horizon probe and therefore such effects should be unsuppressed if they were present during recombination. Vice-versa, tensor modes do affect the polarization of the CMB and therefore the combination of a gravitational slip and propagation speed of GWs can be significantly more informative on the properties of gravity at that early time. Indeed, Ref.~\cite{Amendola:2014wma,Raveri:2014eea} show that the speed of tensors at recombination can be constrained within about $10\%$ of light. The complementarity of probes of background, scalar and tensor sectors has also been explored in Ref.~\cite{Lombriser:2015sxa}.  Other probes of gravitational waves of astrophysical sources have been proposed for constraining modifications of gravity, e.g.\ the decrease of the orbital period of pulsars~\cite{Jimenez:2015bwa}, timing of binary white dwarf systems \cite{Bettoni:2016mij} and the arrival time of GW's compared to that of neutrinos and radiation from supernovae and gamma-ray bursts, respectively~\cite{Nishizawa:2014zna}. Putting all these data together should allow us to answer better what the mechanism for acceleration of the expansion of the universe is.

It is thus our conclusion that dynamical shielding places so onerous a tuning requirement and so strong a dependence on cosmological parameters, that it is impossible to achieve in practice in any natural setting. Thus if future observations were to lead to strong evidence against gravitational slip, they would simultaneously provide a signal that gravitational wave dynamics are standard and thus modifications of gravity that change graviton propagation are unlikely.

\begin{acknowledgments}
We would like to to thank Emilio Bellini and Miguel Zumalacárregui
for helpful comments and discussions. The work of
L.A.~is supported by the DFG through TRR33 ``The Dark Universe''.
M.K.~and M.M.~acknowledge funding by the Swiss National Science
Foundation. IDS is supported by FCT under the grant SFRH/BPD/95204/2013, and further acknowledges UID/FIS/04434/2013 and the project FCT-DAAD 6818/2016-17. I.S. is supported by the European Regional Development Fund and the Czech Ministry of Education, Youth and Sports (MŠMT) (Project CoGraDS -- CZ.02.1.01/0.0/0.0/15\_003/0000437).
\end{acknowledgments}

\bibliographystyle{utcaps}
\bibliography{AnisoRefs,observables,amendola,massive-gravity}

\providecommand{\href}[2]{#2}\begingroup\raggedright\begin{thebibliography}{10}

\bibitem{Amendola:2016saw}
L.~Amendola {\em et~al.}, ``{Cosmology and Fundamental Physics with the Euclid
  Satellite},''
\href{http://arxiv.org/abs/1606.00180}{{\ttfamily arXiv:1606.00180
  [astro-ph.CO]}}.

\bibitem{Amendola:2012ky}
L.~Amendola, M.~Kunz, M.~Motta, I.~D. Saltas, and I.~Sawicki, ``{Observables
  and unobservables in dark energy cosmologies},''
  \href{http://dx.doi.org/10.1103/PhysRevD.87.023501}{{\em Phys.Rev.}
  {\bfseries D87} (2013) 023501},
\href{http://arxiv.org/abs/1210.0439}{{\ttfamily arXiv:1210.0439
  [astro-ph.CO]}}.

\bibitem{Motta:2013cwa}
M.~Motta, I.~Sawicki, I.~D. Saltas, L.~Amendola, and M.~Kunz, ``{Probing Dark
  Energy through Scale Dependence},''
  \href{http://dx.doi.org/10.1103/PhysRevD.88.124035}{{\em Phys.Rev.}
  {\bfseries D88} (2013) 124035},
\href{http://arxiv.org/abs/1305.0008}{{\ttfamily arXiv:1305.0008
  [astro-ph.CO]}}.

\bibitem{Amendola:2013qna}
L.~Amendola, S.~Fogli, A.~Guarnizo, M.~Kunz, and A.~Vollmer,
  ``{Model-independent constraints on the cosmological anisotropic stress},''
\href{http://arxiv.org/abs/1311.4765}{{\ttfamily arXiv:1311.4765
  [astro-ph.CO]}}.

\bibitem{Saltas:2014dha}
I.~D. Saltas, I.~Sawicki, L.~Amendola, and M.~Kunz, ``{Anisotropic Stress as a
  Signature of Nonstandard Propagation of Gravitational Waves},''
  \href{http://dx.doi.org/10.1103/PhysRevLett.113.191101}{{\em Phys.Rev.Lett.}
  {\bfseries 113} no.~19, (2014) 191101},
\href{http://arxiv.org/abs/1406.7139}{{\ttfamily arXiv:1406.7139
  [astro-ph.CO]}}.

\bibitem{Abbott:2016blz}
{\bfseries Virgo, LIGO Scientific} Collaboration, B.~P. Abbott {\em et~al.},
  ``{Observation of Gravitational Waves from a Binary Black Hole Merger},''
  \href{http://dx.doi.org/10.1103/PhysRevLett.116.061102}{{\em Phys. Rev.
  Lett.} {\bfseries 116} no.~6, (2016) 061102},
\href{http://arxiv.org/abs/1602.03837}{{\ttfamily arXiv:1602.03837 [gr-qc]}}.

\bibitem{Horndeski:1974wa}
G.~W. Horndeski, ``{Second-order scalar-tensor field equations in a
  four-dimensional space},''
\href{http://dx.doi.org/10.1007/BF01807638}{{\em Int.J.Theor.Phys.} {\bfseries
  10} (1974) 363--384}.

\bibitem{Deffayet:2011gz}
C.~Deffayet, X.~Gao, D.~Steer, and G.~Zahariade, ``{From k-essence to
  generalised Galileons},''
  \href{http://dx.doi.org/10.1103/PhysRevD.84.064039}{{\em Phys.Rev.}
  {\bfseries D84} (2011) 064039},
\href{http://arxiv.org/abs/1103.3260}{{\ttfamily arXiv:1103.3260 [hep-th]}}.

\bibitem{Jacobson:2000xp}
T.~Jacobson and D.~Mattingly, ``{Gravity with a dynamical preferred frame},''
  \href{http://dx.doi.org/10.1103/PhysRevD.64.024028}{{\em Phys.Rev.}
  {\bfseries D64} (2001) 024028},
\href{http://arxiv.org/abs/gr-qc/0007031}{{\ttfamily arXiv:gr-qc/0007031
  [gr-qc]}}.

\bibitem{Hassan:2011zd}
S.~Hassan and R.~A. Rosen, ``{Bimetric Gravity from Ghost-free Massive
  Gravity},'' \href{http://dx.doi.org/10.1007/JHEP02(2012)126}{{\em JHEP}
  {\bfseries 1202} (2012) 126},
\href{http://arxiv.org/abs/1109.3515}{{\ttfamily arXiv:1109.3515 [hep-th]}}.

\bibitem{Zumalacarregui:2013pma}
M.~Zumalac\'{a}rregui and J.~Garc\'{i}a-Bellido, ``{Transforming gravity: from
  derivative couplings to matter to second-order scalar-tensor theories beyond
  the Horndeski Lagrangian},''
\href{http://arxiv.org/abs/1308.4685}{{\ttfamily arXiv:1308.4685 [gr-qc]}}.

\bibitem{Gleyzes:2014dya}
J.~Gleyzes, D.~Langlois, F.~Piazza, and F.~Vernizzi, ``{Healthy theories beyond
  Horndeski},'' \href{http://dx.doi.org/10.1103/PhysRevLett.114.211101}{{\em
  Phys. Rev. Lett.} {\bfseries 114} no.~21, (2015) 211101},
\href{http://arxiv.org/abs/1404.6495}{{\ttfamily arXiv:1404.6495 [hep-th]}}.

\bibitem{Gleyzes:2014qga}
J.~Gleyzes, D.~Langlois, F.~Piazza, and F.~Vernizzi, ``{Exploring gravitational
  theories beyond Horndeski},''
  \href{http://dx.doi.org/10.1088/1475-7516/2015/02/018}{{\em JCAP} {\bfseries
  1502} (2015) 018},
\href{http://arxiv.org/abs/1408.1952}{{\ttfamily arXiv:1408.1952
  [astro-ph.CO]}}.

\bibitem{Ballesteros:2011cm}
G.~Ballesteros, L.~Hollenstein, R.~K. Jain, and M.~Kunz, ``{Nonlinear
  cosmological consistency relations and effective matter stresses},''
  \href{http://dx.doi.org/10.1088/1475-7516/2012/05/038}{{\em JCAP} {\bfseries
  1205} (2012) 038},
\href{http://arxiv.org/abs/1112.4837}{{\ttfamily arXiv:1112.4837
  [astro-ph.CO]}}.

\bibitem{Adamek:2013wja}
J.~Adamek, D.~Daverio, R.~Durrer, and M.~Kunz, ``{General Relativistic N-body
  simulations in the weak field limit},''
  \href{http://dx.doi.org/10.1103/PhysRevD.88.103527}{{\em Phys.Rev.}
  {\bfseries D88} (2013) 103527},
\href{http://arxiv.org/abs/1308.6524}{{\ttfamily arXiv:1308.6524
  [astro-ph.CO]}}.

\bibitem{Khoury:2003aq}
J.~Khoury and A.~Weltman, ``{Chameleon fields: Awaiting surprises for tests of
  gravity in space},''
  \href{http://dx.doi.org/10.1103/PhysRevLett.93.171104}{{\em Phys.Rev.Lett.}
  {\bfseries 93} (2004) 171104},
\href{http://arxiv.org/abs/astro-ph/0309300}{{\ttfamily arXiv:astro-ph/0309300
  [astro-ph]}}.

\bibitem{Vainshtein:1972sx}
A.~I. Vainshtein, ``{To the problem of nonvanishing gravitation mass},''
\href{http://dx.doi.org/10.1016/0370-2693(72)90147-5}{{\em Phys. Lett.}
  {\bfseries B39} (1972) 393--394}.

\bibitem{Bellini:2014fua}
E.~Bellini and I.~Sawicki, ``{Maximal freedom at minimum cost: linear
  large-scale structure in general modifications of gravity},''
  \href{http://dx.doi.org/10.1088/1475-7516/2014/07/050}{{\em JCAP} {\bfseries
  1407} (2014) 050},
\href{http://arxiv.org/abs/1404.3713}{{\ttfamily arXiv:1404.3713
  [astro-ph.CO]}}.

\bibitem{Gubitosi:2012hu}
G.~Gubitosi, F.~Piazza, and F.~Vernizzi, ``{The Effective Field Theory of Dark
  Energy},'' \href{http://dx.doi.org/10.1088/1475-7516/2013/02/032}{{\em JCAP}
  {\bfseries 1302} (2013) 032},
\href{http://arxiv.org/abs/1210.0201}{{\ttfamily arXiv:1210.0201 [hep-th]}}.

\bibitem{Bloomfield:2012ff}
J.~K. Bloomfield, E.~E. Flanagan, M.~Park, and S.~Watson, ``{Dark energy or
  modified gravity? An effective field theory approach},''
  \href{http://dx.doi.org/10.1088/1475-7516/2013/08/010}{{\em JCAP} {\bfseries
  1308} (2013) 010},
\href{http://arxiv.org/abs/1211.7054}{{\ttfamily arXiv:1211.7054
  [astro-ph.CO]}}.

\bibitem{Gleyzes:2013ooa}
J.~Gleyzes, D.~Langlois, F.~Piazza, and F.~Vernizzi, ``{Essential Building
  Blocks of Dark Energy},''
  \href{http://dx.doi.org/10.1088/1475-7516/2013/08/025}{{\em JCAP} {\bfseries
  1308} (2013) 025},
\href{http://arxiv.org/abs/1304.4840}{{\ttfamily arXiv:1304.4840 [hep-th]}}.

\bibitem{Langlois:2015cwa}
D.~Langlois and K.~Noui, ``{Degenerate higher derivative theories beyond
  Horndeski: evading the Ostrogradski instability},''
  \href{http://dx.doi.org/10.1088/1475-7516/2016/02/034}{{\em JCAP} {\bfseries
  1602} no.~02, (2016) 034},
\href{http://arxiv.org/abs/1510.06930}{{\ttfamily arXiv:1510.06930 [gr-qc]}}.

\bibitem{Crisostomi:2016tcp}
M.~Crisostomi, M.~Hull, K.~Koyama, and G.~Tasinato, ``{Horndeski: beyond, or
  not beyond?},'' \href{http://dx.doi.org/10.1088/1475-7516/2016/03/038}{{\em
  JCAP} {\bfseries 1603} no.~03, (2016) 038},
\href{http://arxiv.org/abs/1601.04658}{{\ttfamily arXiv:1601.04658 [hep-th]}}.

\bibitem{Bonvin:2007ap}
C.~Bonvin, R.~Durrer, P.~G. Ferreira, G.~Starkman, and T.~G. Zlosnik,
  ``{Generalized Einstein-Aether theories and the Solar System},''
  \href{http://dx.doi.org/10.1103/PhysRevD.77.024037}{{\em Phys. Rev.}
  {\bfseries D77} (2008) 024037},
\href{http://arxiv.org/abs/0707.3519}{{\ttfamily arXiv:0707.3519 [astro-ph]}}.

\bibitem{Yagi:2013ava}
K.~Yagi, D.~Blas, E.~Barausse, and N.~Yunes, ``{Constraints on Einstein-\AE
  ther theory and Horava gravity from binary pulsar observations},''
  \href{http://dx.doi.org/10.1103/PhysRevD.89.084067}{{\em Phys.Rev.}
  {\bfseries D89} (2014) 084067},
\href{http://arxiv.org/abs/1311.7144}{{\ttfamily arXiv:1311.7144 [gr-qc]}}.

\bibitem{Yagi:2013qpa}
K.~Yagi, D.~Blas, N.~Yunes, and E.~Barausse, ``{Strong Binary Pulsar
  Constraints on Lorentz Violation in Gravity},''
  \href{http://dx.doi.org/10.1103/PhysRevLett.112.161101}{{\em Phys. Rev.
  Lett.} {\bfseries 112} no.~16, (2014) 161101},
\href{http://arxiv.org/abs/1307.6219}{{\ttfamily arXiv:1307.6219 [gr-qc]}}.

\bibitem{Lim:2004js}
E.~A. Lim, ``{Can we see Lorentz-violating vector fields in the CMB?},''
  \href{http://dx.doi.org/10.1103/PhysRevD.71.063504}{{\em Phys.Rev.}
  {\bfseries D71} (2005) 063504},
\href{http://arxiv.org/abs/astro-ph/0407437}{{\ttfamily arXiv:astro-ph/0407437
  [astro-ph]}}.

\bibitem{Jacobson:2013xta}
T.~Jacobson, ``{Undoing the twist: the Ho\v{r}ava limit of Einstein-aether},''
\href{http://arxiv.org/abs/1310.5115}{{\ttfamily arXiv:1310.5115 [gr-qc]}}.

\bibitem{deRham:2010ik}
C.~de~Rham and G.~Gabadadze, ``{Generalization of the Fierz-Pauli Action},''
  \href{http://dx.doi.org/10.1103/PhysRevD.82.044020}{{\em Phys.Rev.}
  {\bfseries D82} (2010) 044020},
\href{http://arxiv.org/abs/1007.0443}{{\ttfamily arXiv:1007.0443 [hep-th]}}.

\bibitem{deRham:2011rn}
C.~de~Rham, G.~Gabadadze, and A.~J. Tolley, ``{Ghost free Massive Gravity in
  the St\'uckelberg language},''
  \href{http://dx.doi.org/10.1016/j.physletb.2012.03.081}{{\em Phys.Lett.}
  {\bfseries B711} (2012) 190--195},
\href{http://arxiv.org/abs/1107.3820}{{\ttfamily arXiv:1107.3820 [hep-th]}}.

\bibitem{deRham:2014zqa}
C.~de~Rham, ``{Massive Gravity},''
  \href{http://dx.doi.org/10.12942/lrr-2014-7}{{\em Living Rev.Rel.} {\bfseries
  17} (2014) 7},
\href{http://arxiv.org/abs/1401.4173}{{\ttfamily arXiv:1401.4173 [hep-th]}}.

\bibitem{Schmidt-May:2015vnx}
A.~Schmidt-May and M.~von Strauss, ``{Recent developments in bimetric
  theory},'' \href{http://dx.doi.org/10.1088/1751-8113/49/18/183001}{{\em J.
  Phys.} {\bfseries A49} no.~18, (2016) 183001},
\href{http://arxiv.org/abs/1512.00021}{{\ttfamily arXiv:1512.00021 [hep-th]}}.

\bibitem{deRham:2014fha}
C.~de~Rham, L.~Heisenberg, and R.~H. Ribeiro, ``{Ghosts and matter couplings in
  massive gravity, bigravity and multigravity},''
  \href{http://dx.doi.org/10.1103/PhysRevD.90.124042}{{\em Phys. Rev.}
  {\bfseries D90} (2014) 124042},
\href{http://arxiv.org/abs/1409.3834}{{\ttfamily arXiv:1409.3834 [hep-th]}}.

\bibitem{Comelli:2012db}
D.~Comelli, M.~Crisostomi, and L.~Pilo, ``{Perturbations in Massive Gravity
  Cosmology},'' \href{http://dx.doi.org/10.1007/JHEP06(2012)085}{{\em JHEP}
  {\bfseries 1206} (2012) 085},
\href{http://arxiv.org/abs/1202.1986}{{\ttfamily arXiv:1202.1986 [hep-th]}}.

\bibitem{Solomon:2014dua}
A.~R. Solomon, Y.~Akrami, and T.~S. Koivisto, ``{Linear growth of structure in
  massive bigravity},''
  \href{http://dx.doi.org/10.1088/1475-7516/2014/10/066}{{\em JCAP} {\bfseries
  1410} (2014) 066},
\href{http://arxiv.org/abs/1404.4061}{{\ttfamily arXiv:1404.4061
  [astro-ph.CO]}}.

\bibitem{Cusin:2014psa}
G.~Cusin, R.~Durrer, P.~Guarato, and M.~Motta, ``{Gravitational waves in
  bigravity cosmology},''
  \href{http://dx.doi.org/10.1088/1475-7516/2015/05/030}{{\em JCAP} {\bfseries
  1505} no.~05, (2015) 030},
\href{http://arxiv.org/abs/1412.5979}{{\ttfamily arXiv:1412.5979
  [astro-ph.CO]}}.

\bibitem{Linder:2014fna}
E.~V. Linder, ``{Are Scalar and Tensor Deviations Related in Modified
  Gravity?},'' \href{http://dx.doi.org/10.1103/PhysRevD.90.083536}{{\em
  Phys.Rev.} {\bfseries D90} (2014) 083536},
\href{http://arxiv.org/abs/1407.8184}{{\ttfamily arXiv:1407.8184
  [astro-ph.CO]}}.

\bibitem{Linder:2015rcz}
E.~V. Linder, G.~Sengör, and S.~Watson, ``{Is the Effective Field Theory of
  Dark Energy Effective?},''
  \href{http://dx.doi.org/10.1088/1475-7516/2016/05/053}{{\em JCAP} {\bfseries
  1605} no.~05, (2016) 053},
\href{http://arxiv.org/abs/1512.06180}{{\ttfamily arXiv:1512.06180
  [astro-ph.CO]}}.

\bibitem{Bellini:2013hea}
E.~Bellini and R.~Jimenez, ``{The parameter space of Cubic Galileon models for
  cosmic acceleration},''
  \href{http://dx.doi.org/10.1016/j.dark.2013.11.001}{{\em Phys.Dark Univ.}
  {\bfseries 2} (2013) 179--183},
\href{http://arxiv.org/abs/1306.1262}{{\ttfamily arXiv:1306.1262
  [astro-ph.CO]}}.

\bibitem{Blas:2016qmn}
D.~Blas, M.~M. Ivanov, I.~Sawicki, and S.~Sibiryakov, ``{On constraining the
  speed of gravitational waves following GW150914},''
  \href{http://dx.doi.org/10.1134/S0021364016100040,
  10.7868/S0370274X16100039}{{\em Pisma Zh. Eksp. Teor. Fiz.} {\bfseries 103}
  no.~10, (2016) 708--710}, \href{http://arxiv.org/abs/1602.04188}{{\ttfamily
  arXiv:1602.04188 [gr-qc]}}.
[JETP Lett.103,no.10,624(2016)].

\bibitem{Sawicki:2015zya}
I.~Sawicki and E.~Bellini, ``{Limits of quasistatic approximation in
  modified-gravity cosmologies},''
  \href{http://dx.doi.org/10.1103/PhysRevD.92.084061}{{\em Phys. Rev.}
  {\bfseries D92} no.~8, (2015) 084061},
\href{http://arxiv.org/abs/1503.06831}{{\ttfamily arXiv:1503.06831
  [astro-ph.CO]}}.

\bibitem{Lombriser:2014ira}
L.~Lombriser and A.~Taylor, ``{Classifying Linearly Shielded Modified Gravity
  Models in Effective Field Theory},''
  \href{http://dx.doi.org/10.1103/PhysRevLett.114.031101}{{\em Phys. Rev.
  Lett.} {\bfseries 114} no.~3, (2015) 031101},
\href{http://arxiv.org/abs/1405.2896}{{\ttfamily arXiv:1405.2896
  [astro-ph.CO]}}.

\bibitem{Lombriser:2015sxa}
L.~Lombriser and A.~Taylor, ``{Breaking a Dark Degeneracy with Gravitational
  Waves},'' \href{http://dx.doi.org/10.1088/1475-7516/2016/03/031}{{\em JCAP}
  {\bfseries 1603} no.~03, (2016) 031},
\href{http://arxiv.org/abs/1509.08458}{{\ttfamily arXiv:1509.08458
  [astro-ph.CO]}}.

\bibitem{Konnig:2013gxa}
F.~Koennig, A.~Patil, and L.~Amendola, ``{Viable cosmological solutions in
  massive bimetric gravity},''
  \href{http://dx.doi.org/10.1088/1475-7516/2014/03/029}{{\em JCAP} {\bfseries
  1403} (2014) 029},
\href{http://arxiv.org/abs/1312.3208}{{\ttfamily arXiv:1312.3208
  [astro-ph.CO]}}.

\bibitem{Konnig:2014xva}
F.~Koennig, Y.~Akrami, L.~Amendola, M.~Motta, and A.~R. Solomon, ``{Stable and
  unstable cosmological models in bimetric massive gravity},''
  \href{http://dx.doi.org/10.1103/PhysRevD.90.124014}{{\em Phys. Rev.}
  {\bfseries D90} (2014) 124014},
\href{http://arxiv.org/abs/1407.4331}{{\ttfamily arXiv:1407.4331
  [astro-ph.CO]}}.

\bibitem{Lagos:2014lca}
M.~Lagos and P.~G. Ferreira, ``{Cosmological perturbations in massive
  bigravity},'' \href{http://dx.doi.org/10.1088/1475-7516/2014/12/026}{{\em
  JCAP} {\bfseries 1412} (2014) 026},
\href{http://arxiv.org/abs/1410.0207}{{\ttfamily arXiv:1410.0207 [gr-qc]}}.

\bibitem{Alonso:2015sfa}
D.~Alonso and P.~G. Ferreira, ``{Constraining ultralarge-scale cosmology with
  multiple tracers in optical and radio surveys},''
  \href{http://dx.doi.org/10.1103/PhysRevD.92.063525}{{\em Phys. Rev.}
  {\bfseries D92} no.~6, (2015) 063525},
\href{http://arxiv.org/abs/1507.03550}{{\ttfamily arXiv:1507.03550
  [astro-ph.CO]}}.

\bibitem{Pogosian:2016pwr}
L.~Pogosian and A.~Silvestri, ``{What can cosmology tell us about gravity?
  Constraining Horndeski gravity with $\Sigma$ and $\mu$},''
  \href{http://dx.doi.org/10.1103/PhysRevD.94.104014}{{\em Phys. Rev.}
  {\bfseries D94} no.~10, (2016) 104014},
\href{http://arxiv.org/abs/1606.05339}{{\ttfamily arXiv:1606.05339
  [astro-ph.CO]}}.

\bibitem{Amendola:2014wma}
L.~Amendola, G.~Ballesteros, and V.~Pettorino, ``{Effects of modified gravity
  on B-mode polarization},''
  \href{http://dx.doi.org/10.1103/PhysRevD.90.043009}{{\em Phys. Rev.}
  {\bfseries D90} (2014) 043009},
\href{http://arxiv.org/abs/1405.7004}{{\ttfamily arXiv:1405.7004
  [astro-ph.CO]}}.

\bibitem{Raveri:2014eea}
M.~Raveri, C.~Baccigalupi, A.~Silvestri, and S.-Y. Zhou, ``{Measuring the speed
  of cosmological gravitational waves},''
  \href{http://dx.doi.org/10.1103/PhysRevD.91.061501}{{\em Phys. Rev.}
  {\bfseries D91} no.~6, (2015) 061501},
\href{http://arxiv.org/abs/1405.7974}{{\ttfamily arXiv:1405.7974
  [astro-ph.CO]}}.

\bibitem{Jimenez:2015bwa}
J.~Beltran~Jimenez, F.~Piazza, and H.~Velten, ``{Evading the Vainshtein
  Mechanism with Anomalous Gravitational Wave Speed: Constraints on Modified
  Gravity from Binary Pulsars},''
  \href{http://dx.doi.org/10.1103/PhysRevLett.116.061101}{{\em Phys. Rev.
  Lett.} {\bfseries 116} no.~6, (2016) 061101},
\href{http://arxiv.org/abs/1507.05047}{{\ttfamily arXiv:1507.05047 [gr-qc]}}.

\bibitem{Bettoni:2016mij}
D.~Bettoni, J.~M. Ezquiaga, K.~Hinterbichler, and M.~Zumalacárregui,
  ``{Gravitational Waves and the Fate of Scalar-Tensor Gravity},''
\href{http://arxiv.org/abs/1608.01982}{{\ttfamily arXiv:1608.01982 [gr-qc]}}.

\bibitem{Nishizawa:2014zna}
A.~Nishizawa and T.~Nakamura, ``{Measuring Speed of Gravitational Waves by
  Observations of Photons and Neutrinos from Compact Binary Mergers and
  Supernovae},'' \href{http://dx.doi.org/10.1103/PhysRevD.90.044048}{{\em Phys.
  Rev.} {\bfseries D90} no.~4, (2014) 044048},
\href{http://arxiv.org/abs/1406.5544}{{\ttfamily arXiv:1406.5544 [gr-qc]}}.

\end{thebibliography}\endgroup

\end{document}